\documentclass[%
 reprint,
superscriptaddress,
 amsmath,amssymb,
 aps,
pra,
showpacs,
]{revtex4-1}

\usepackage{graphicx}
\usepackage{dcolumn}
\usepackage{bm}
\usepackage{graphicx}
\usepackage{epstopdf}
\usepackage{subfigure}

\begin{document}

\title{The interplay between resonant enhancement and quantum path dynamics in harmonic generation in helium}

\author{Seth Camp }
\email{scamp1@lsu.edu}
\author{Kenneth J. Schafer}%
\author{Mette B. Gaarde}
\email{gaarde@phys.lsu.edu}
\affiliation{%
 Deparment of Physics and Astronomy, Louisiana State University, Baton Rouge, Louisiana 70803-4001
}%

\date{\today}

\begin{abstract}
We present a theoretical study of the influence of resonant enhancement on quantum path dynamics in the generation of harmonics above and below the ionization threshold in helium. By varying the wavelength and intensity of the driving field from 425 nm to 500 nm and from 30 TW/cm${^2}$ to 140 TW/cm${^2}$, respectively, we identify enhancements of harmonics 7, 9, and 11 that correspond to multiphoton resonances between the ground state and the Stark shifted $1s2p$, $1s3p$, and $1s4p$ excited states. A time-frequency analysis of the emission shows that both the short and long quantum path contributions to the harmonic yield are enhanced through these bound state resonances. We analyze the sub-cycle time structure of the 9th harmonic yield in the vicinity of the resonances and find that on resonance the long trajectory contribution is phase shifted by approximately $\pi/4$.  Finally, we compare the single atom and the macroscopic response of a helium gas and find that while the sub-cycle time profiles are slightly distorted by propagation effects,  the phase shift of the long trajectory contribution is still recognizable. 
\end{abstract}

\maketitle

\section{Introduction}

High harmonic generation is a versatile source of ultrafast, coherent extreme ultraviolet (XUV) radiation \cite{McPherson-1987,Ferray-1988}, produced by the interaction between an intense infrared or visible laser field and a gas of atoms or molecules. The semi-classical understanding of harmonic generation, in which an electron wave packet is initiated by tunnel ionization, accelerates in the laser field, and finally re-scatters on the parent ion \cite{Schafer-1993, Corkum-1994} has given rise to the field of high harmonic spectroscopy (HHS), in which the rescattering electron wave packet has been used as a sensitive probe of the structure and dynamics of the ion core \cite{Itatani-2004,Morishita-2008,Le-2008,Smirnova-2009,Worner-2010}. 

Resonantly enhanced high harmonic generation (REHHG) is a particular example of the more general HHS technique. Although REHHG has been shown to increase the harmonic yield in a limited range of experimental settings \cite{Toma-1999, Schafer-2001, Figueira-2002, Taieb-2003, Strelkov-2010, Ackermann-2013}, it has more recently been explored for its potential to learn about the  dynamics of bound and quasi-bound states in the presence of a strong driving field \cite{Strelkov-2010, Redkin-2011, Shiner-2011, Lein-2011, Jin-2012, Chu-2013, Gong-2013, Chini-2014, Ferre-2014}. Several mechanisms for resonant enhancement have been discussed in the literature, generally all involving an intermediate, resonant, step in the semi-classical model. The resonant step may occur either in the ionization process, via a multiphoton resonance between the ground state and the Stark-shifted excited state, or in the rescattering process via enhanced recombination, or by capture into an excited bound state that subsequently decays via spontaneous emission of light  \cite{Taieb-2003, Strelkov-2010, Redkin-2011, Lein-2011, Chu-2013, Gong-2013}. The capture and spontaneous emission process has been explored in detail for short-lived quasi-bound states embedded in a continuum for which it can give rise to very large enhancements \cite{Strelkov-2010, Lein-2011}. For bound-state resonances with long lifetimes, the capture and spontaneous emission process can generally be distinguished from the coherently driven resonant enhancement (via multiphoton ionization or  enhanced recombination), since it will give rise to narrow-band radiation at the field-free resonance frequency given that it largely takes place after the driving laser pulse is over \cite{Chini-2014, Li-2014}. In contrast to this, the coherently driven resonantly enhanced response will give rise to emission at the difference frequency between the ground state and the Stark-shifted excited state since this process only takes place while the laser field is on.

In this paper we study the coherently driven REHHG and investigate the interplay between the resonant enhancement and the quantum path dynamics of the harmonic generation process. In particular, we study how the amplitude and phase of the different quantum path contributions to the harmonic yield in helium are changed in the vicinity of a bound state resonance. We do this by first solving the time-dependent Schr{\"o}dinger equation (TDSE) in the single active electron (SAE) approximation for wavelengths between $425-500$ nm and for intensities up to $140$ TW/cm$^2$. We have chosen the intensity and wavelength regime such that the Keldysh parameter, of approximately 2, is in line with previous studies \cite{Schafer-2001, Ackermann-2013, Gong-2013}. We identify resonant enhancement of harmonics 7, 9, and 11 (H7, H9, H11) due to Stark-shifted resonances between the ground state and the $1s2p$, $1s3p$, and $1s4p$ states. By considering the time-frequency profiles of the harmonic emission, we also clearly identify the familiar short and long quantum path contributions to these harmonics both above and below the field-free ionization threshold. We find that close to the $1s3p$ resonance, the amplitude of both the short and long quantum path contributions are enhanced. We also find that  the long-trajectory contribution  is phase shifted by approximately $\pi/4$ in the vicinity of this resonance, while the short-trajectory contribution is not. These results indicate that while the enhancement of the yield happens in the ionization step, the phase shift is due to the interaction between the returning electron wave packet and the atomic potential, for which there is a large difference between the short and long trajectory dynamics. Finally, we  calculate the macroscopic harmonic response by solving the coupled Maxwell wave equation and the TDSE for a gas of atoms interacting with a focused laser beam. We find that the enhancement and the phase shift of the resonantly enhanced harmonics can still be recognized in the macroscopic response. 

The paper is structured as follows: Section II presents a brief introduction to the theoretical methods used in this paper. Section III and IV focus on the single atom response; in Section III we look at REHHG in the frequency domain and use the harmonic spectra to map out the enhancement as a function of driving intensity and wavelength, and in Section IV we examine REHHG in the time domain and study the effect of REHHG on the quantum trajectories. Section V presents results of our macroscopic calculations, and Section IV presents a summary of our findings. 


\section{Theoretical Method}

We numerically solve the TDSE in the SAE approximation for a He atom interacting with an intense laser field. We use a pseudo-potential that  reproduces well the energies and oscillator strengths of the singly excited states of helium to describe the interaction of the active electron with the field. We will generally refer to the excited $1snp$ states as $np$. For more details on our numerical method and the pseudo-potential see \cite{Schafer-Strongfield}. The laser field has the form of a cosine squared pulse with a full width at half maximum (FWHM) duration $\tau$, given by:
\begin{equation}
E_{L}(t) = E_0 \cos(\frac{c_2 t}{\tau})^2  \cos(2\pi t),
\end{equation}
where $E_0$ is the peak electric field, $c_2 = 2\arccos(\frac{1}{2}^{\frac{1}{4}})$,  and $t$ is  time in units of optical cycles (O.C.) of the driving field \cite{Barth-2009}. With a  FWHM duration $\tau = 6$ O.C. the total duration of the pulse, and thereby the TDSE integration, is 16.5 O.C. We calculate the dipole spectrum from the time-dependent acceleration $a(t)$ as $\widetilde{D}(\omega) = -\widetilde{A}(\omega)/\omega^2$, where the acceleration spectrum is given by: 
\begin{equation}
\widetilde{A}(\omega) = \int a(t) W_H(t) e^{i \omega t} dt.
\end{equation}
In this equation, $W_H(t) = .5(1+\cos(\frac{\pi t}{\tau_H}))$ is a Hanning window function \cite{Blackman-1958} which brings the time-dependent acceleration smoothly to zero at the end of the calculation. In numerical calculations it is always necessary to include a window function on $a(t)$ when there is any population remaining in excited states at the end of the pulse, since the coherence between the excited and the ground state population gives rise to a dipole moment which in the calculation would last forever. In an experiment, this would give rise to spontaneous emission, decaying with the life time of the excited states. The window function thus serves to impose an artificial lifetime on the excited states. In most of the calculations shown in this paper, we use a Hanning window function with $\tau_H = 8.25$ O.C. in order to last only as long as the driving laser pulse. This means we are primarily considering the coherent, driven, response of the atom to the laser pulse and not the radiation produced by long-lived decay of excited state populations. It also means that in most of the calculations shown, we are discarding the capture and spontaneous emission contribution to the resonantly enhanced response. By doing a few calculations where we continue the TDSE integration after the end of the laser pulse (using a longer Hanning window where $\tau_H=8.25$ O.C. for $t<0$ and $\tau_H=30$ O.C. for $t>0$) we will show that this latter contribution gives rise to narrow peaks at the position of the field-free resonant frequencies which, at the intensities used in this study, is generally outside or in the wings of the harmonic spectral bandwidths.  

We also calculate the macroscopic harmonic response by solving the  coupled  Maxwell wave equation and  TDSE (MWE-TDSE) via space-marching of the full electric field $\widetilde{E}(\omega)$, containing both laser and harmonics fields, through the nonlinear medium \cite{Gaarde-2011}. In the slowly evolving wave approximation, we can express the MWE in the following form (in SI units): 
\begin{equation}
\nabla^2_\perp \widetilde{E}(\omega) + \frac{2 i \omega}{c} \frac{\partial}{\partial z} \widetilde{E}(\omega) = -\frac{\omega^2}{\epsilon_0 c^2}[ \widetilde{P}(\omega)+\widetilde{P}_{ion}(\omega)],
\label{MWE}
\end{equation}
where $\widetilde{E}(\omega)$, $\widetilde{P}(\omega)$, and $\widetilde{P}_{ion}(\omega)$ are also functions of the cylindrical coordinates $r$ and $z$. $\widetilde{P}(\omega)=2N_{at}\widetilde{D}(\omega)$ is the macroscopic polarization field, where $N_{at}$ is the atomic density and $\widetilde{D}(\omega)$ is the one-electron single-atom dipole moment. $\widetilde{P}_{ion}$ is related to the free-electron contribution to the refractive index and is ignored for these calculations as it is a small term for the parameters we are studying. At each step in the propagation direction $z$, we calculate $\widetilde{P}(\omega)$  in Eq.~(\ref{MWE}) by solving the TDSE at each radial point, and then propagate Eq.~(\ref{MWE}) one step to get to the next $z$ plane. More details of our MWE-TDSE solution can be found in \cite{Gaarde-2011, Mette-Chapter2}.

In the macroscopic calculations, the focused laser beam is modeled as a Gaussian beam with a peak intensity of $140$ TW/cm$^2$, a confocal parameter of 1 cm, and the focus of the beam is 1 mm before the center of the 1 mm long gas medium that has a pressure of approximately $30$ Torr. We use the same temporal profile as in the single atom calculations. All macroscopic calculations shown in this paper use the radially integrated yield of the electric field intensity at the end of the medium. 


\section{Single Atom: Frequency Domain.}

\begin{figure}[t!]
\scalebox{.7}{
		\includegraphics[trim = 8mm 0mm 0mm 0mm,clip]{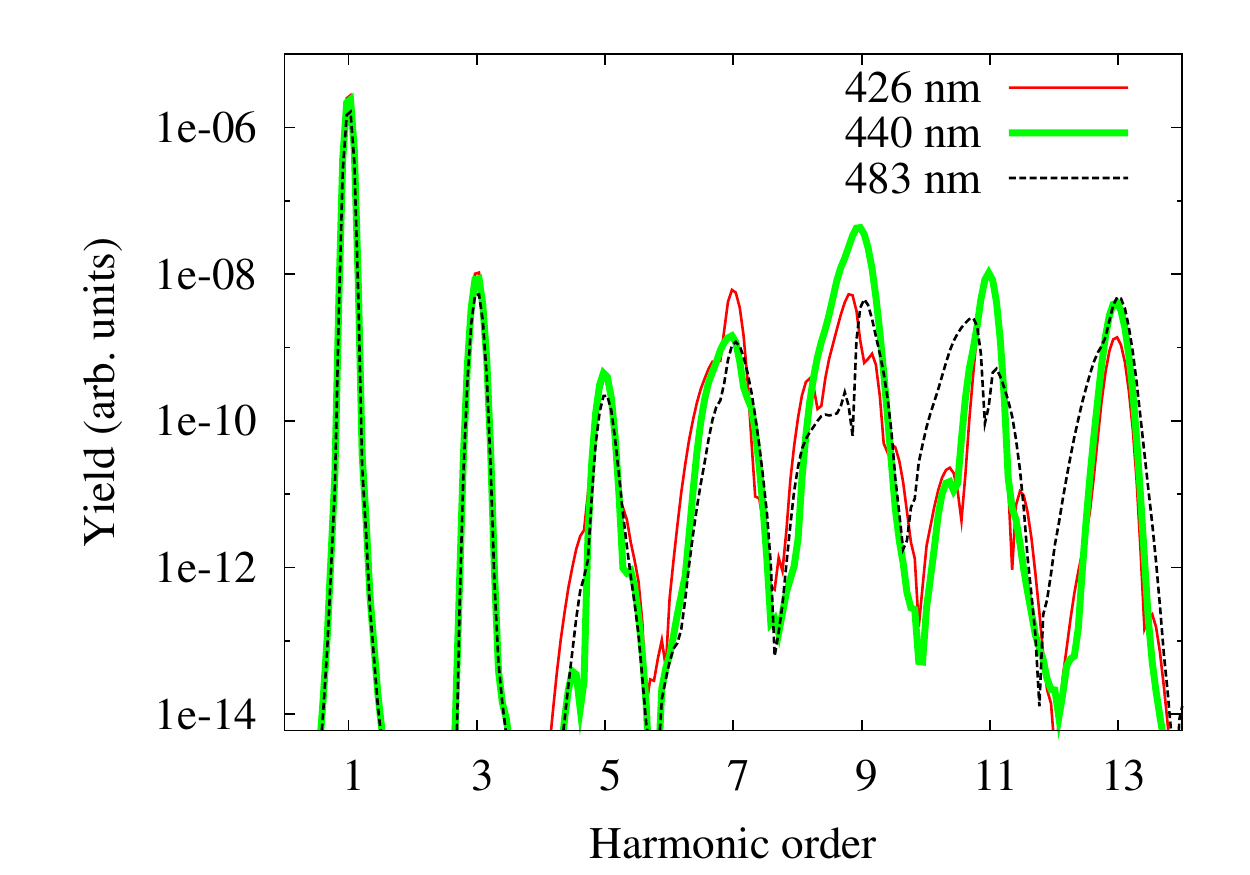}}
		\caption{Harmonic spectra for different driving wavelengths for a driving laser intensity of 140 TW/cm$^2$.}
	\label{fig:sa-spectra}
\end{figure}

In this section we characterize the resonant enhancements in the harmonic spectrum as a function of laser wavelength and intensity. By "resonant enhancement" we mean that, for a given intensity, a harmonic is much stronger at a particular wavelength than at others because the harmonic energy is  resonant with a transition between the ground state and a Stark-shifted excited state \cite{Freeman-1987, Agostini-1989-2}. For most of the $np$-states, the Stark shift is approximately equal to the ponderomotive shift of the continuum states \cite{Ackermann-2013, Schafer-2001,Figueira-2002}. Thus direct resonant enhancement of harmonic $q$ occurs when the following equation is satisfied : 
\begin{equation}
|E_{np} - E_0| +U_p = q \hbar \omega,
\end{equation}
where $E_{np}$  and $E_0$ are the field-free energies of the $np$-state and the ground state of helium, respectively, $U_p$ is the ponderomotive energy, and $\hbar \omega$ is the laser photon energy. Indirect resonant enhancements occur when Eq. (4) is satisfied and a harmonic $q'$ near harmonic $q$ is enhanced. 

\begin{figure}[t]
		\includegraphics[width=8.5cm, trim = 42mm 4mm 22mm 4mm, clip]{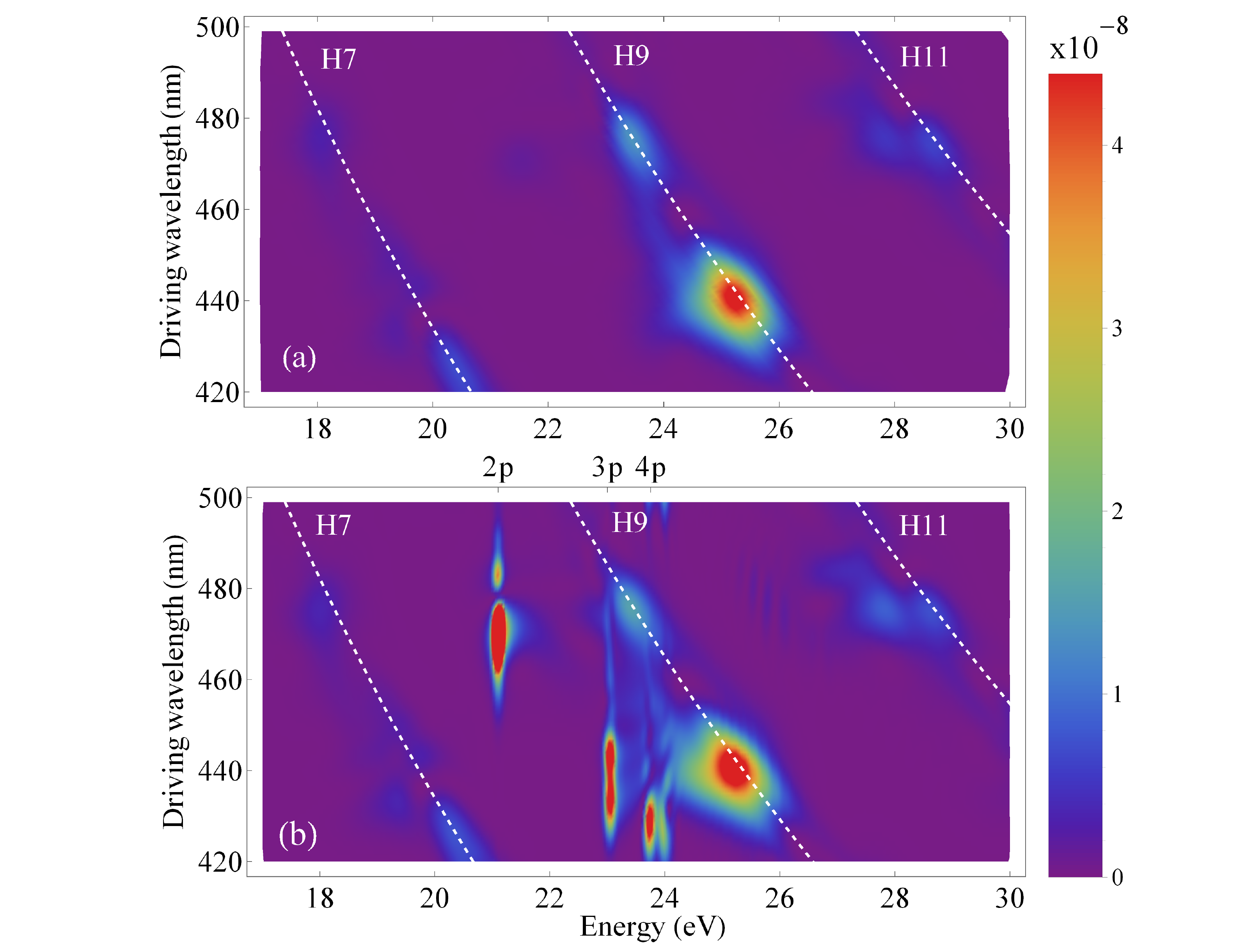}
		\caption{Comparison of the harmonic spectra as functions of driving wavelength and energy for different Hanning windows for a driving intensity of $140$ TW/cm$^2$. (a) shows the spectra for a Hanning window that matches the duration of the pulse. (b) shows the spectra for a Hanning window that allows the dipole to ring for approximately 20 optical cycles after the pulse ends. The white dotted lines mark the position of H7-H11 as a function of driving wavelength. In (b) we have also indicated the field-free energies of the $2p$-$5p$ states. We note that the harmonic yield is shown on a linear scale, in arbitrary units.}
		\label{fig:ringing-spectra}
\end{figure}

Fig. \ref{fig:sa-spectra} shows  harmonic spectra for driving wavelengths of $426$, $440$, and $483$ nm at an  intensity of $140$ TW/cm$^2$. We see two enhancements in the spectra, H9 at $440$ nm and H7 at $426$ nm. Eq. (4) shows that for these three wavelengths, it is only the enhancement of H9 at 440 nm which represents a direct enhancement, via a 9-photon resonance between the ground state and the Stark-shifted $3p$ state. As we will show below, the enhancement of H7 at 426 nm is an indirect enhancement, via a 9-photon resonance between the ground state and the $4p$ state. 

Fig. \ref{fig:ringing-spectra} shows the spectral yield for a range of driving wavelengths between 420 nm and 500 nm. We have used two different Hanning windows to be able to distinguish between the coherently driven response, and the response due to the capture and spontaneous emission process. The shorter window in Fig. \ref{fig:ringing-spectra}(a) shows the coherently driven response that we are predominantly interested in, whereas the longer window used in Fig. \ref{fig:ringing-spectra}(b) additionally shows the radiation due to population left in excited states at the end of the pulse. The narrow spectral lines that are visible at the field-free transition frequencies in Fig. \ref{fig:ringing-spectra}(b) are useful in identifying which state is primarily responsible for a given resonant enhancement. For example, the enhancement of H9 at 440 nm (470 nm) is clearly associated with a large population transfer to the $3p$ ($2p$) state whereas the modest enhancement of H7 at 426 nm is associated with 9-photon-driven population transfer to the $4p$ state. Interestingly, this $4p$ resonance does not lead to a particular enhancement of H9, at least in comparison with the large enhancement due to the $3p$ state.  It is worth noting the large spectral separation between the coherently enhanced radiation and the field-free peaks due to the long-lived dipole moment.

\begin{figure}[t]
	\centering
	 \scalebox{.36}{
		\includegraphics{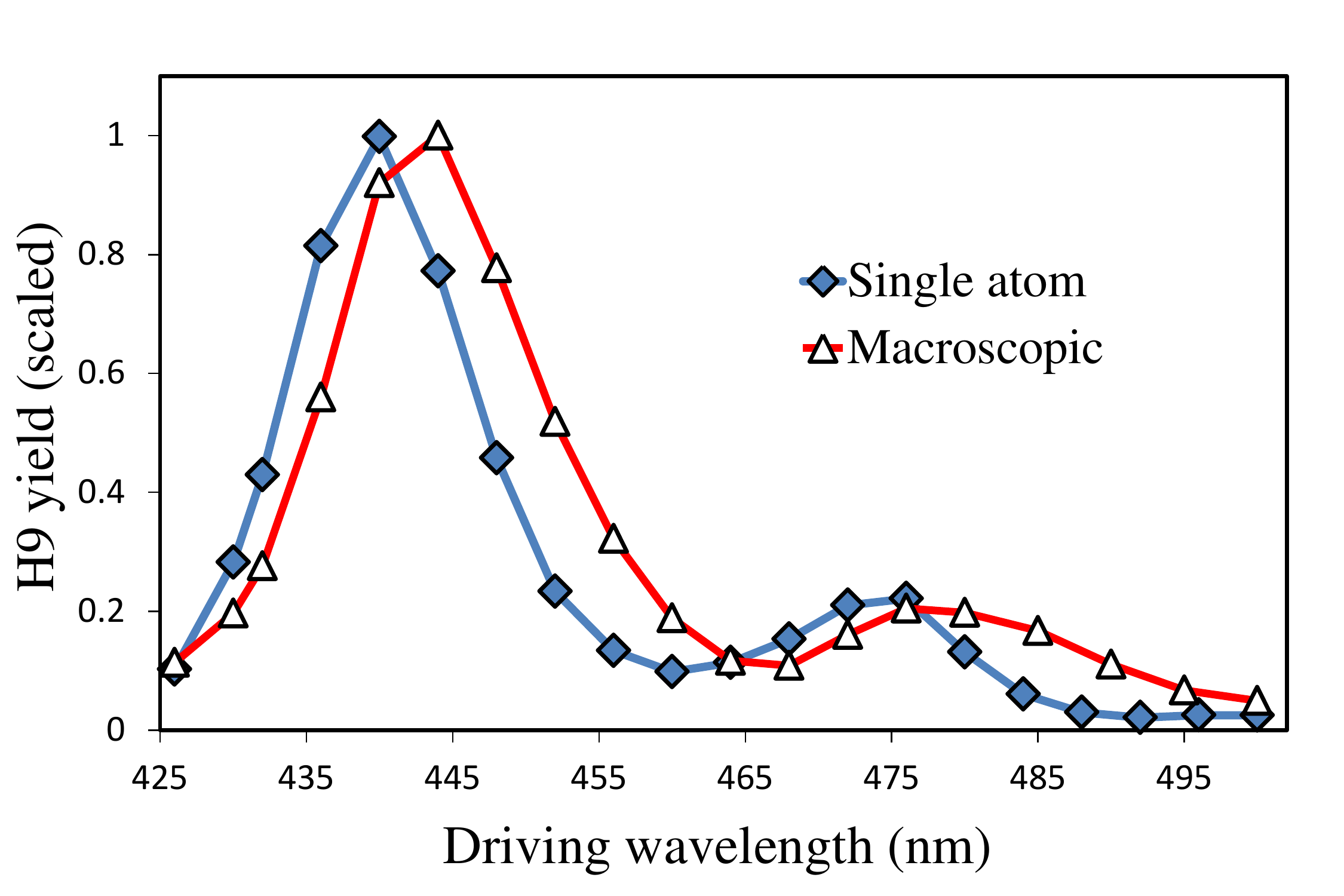}}
		\caption{H9 spectrally integrated yield as a function of driving wavelength, for an intensity of 140 TW/cm$^2$. We show both the single atom yield  (solid diamonds) and the macroscopic yield (open triangles). We have used the shorter Hanning window in calculating the harmonic spectrum.}
	\label{fig:SA_vs_Macro}	 
\end{figure}

The resonant enhancements in any given harmonic can also be followed by considering the combined wavelength and intensity dependence of that harmonic. We start by showing in Fig. \ref{fig:SA_vs_Macro} the wavelength dependence of the H9 yield, using an intensity of $140$ TW/cm$^2$.  This plot corresponds to a line-out of Fig.~\ref{fig:ringing-spectra}(a) along the white dashed line indicating the central frequency of H9, spectrally integrated from H8 to H10. At this intensity, one can recognize two enhancement features in Fig. \ref{fig:SA_vs_Macro}, a strong enhancement centered at $440$ nm and a weaker enhancement centered at $476$ nm, as we would expect from Fig. \ref{fig:ringing-spectra}. The location of these peaks agree well with the prediction of Eq. (4) for the $3p$ and $2p$ states and $q=9$. 

\begin{figure}[t]
	\centering
	 \scalebox{.35}{
		\includegraphics{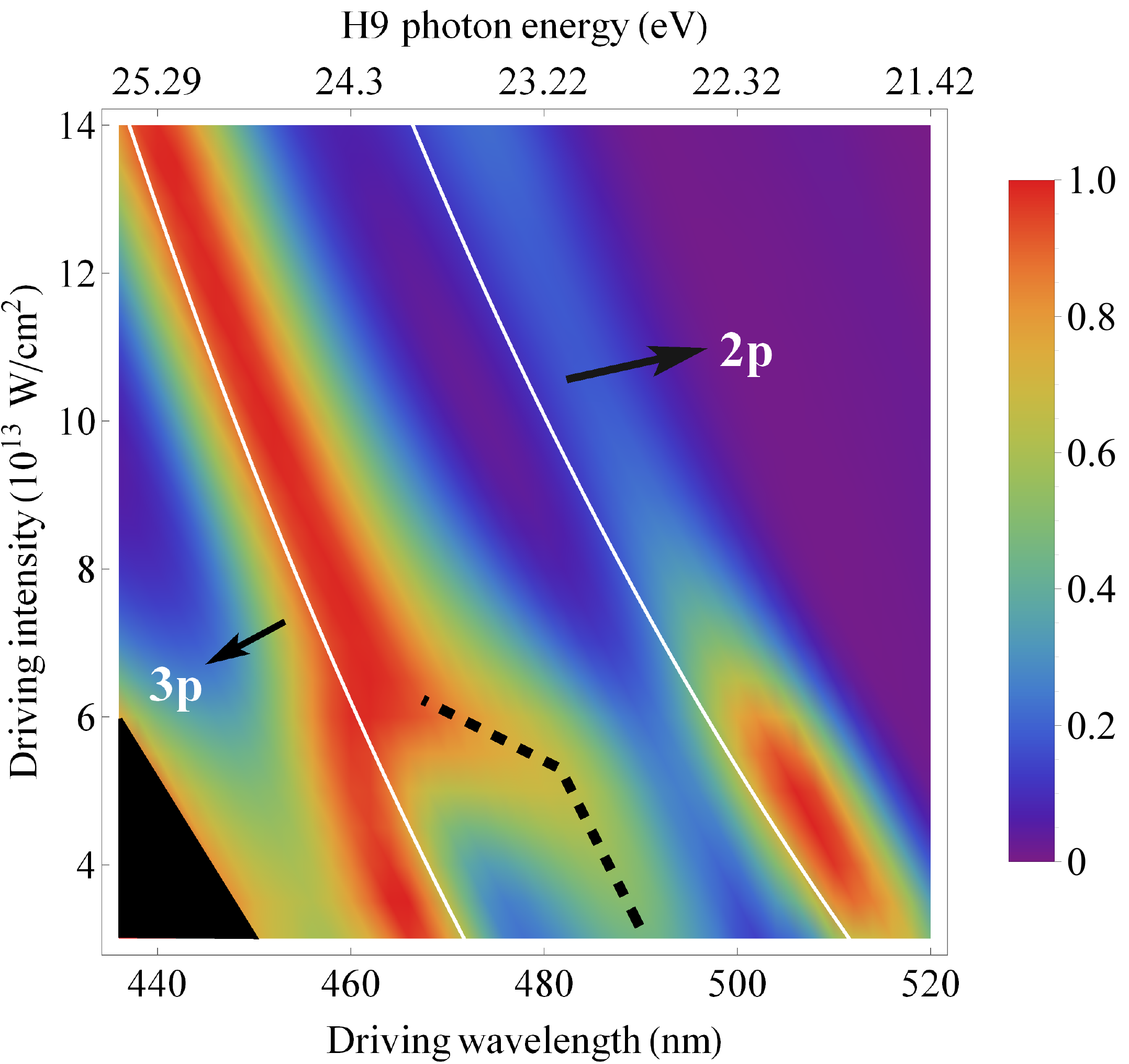}}
		\caption{H9 spectrally integrated yield as a function  of driving wavelength and intensity. For each intensity, the spectrum has been normalized. The white lines represent the  photon energies of the resonant enhancements due to the $2p$ and $3p$ states predicted by Eq. (4). The dark dotted line highlights the lower energy feature of an Autler-Townes splitting of the $3p$ feature (see text).}
	\label{fig:EnhancementMap}
\end{figure}

Next, Fig.~\ref{fig:EnhancementMap} shows the combined intensity and wavelength dependence of H9, by plotting the H9 wavelength dependence as shown in Fig. \ref{fig:SA_vs_Macro} for many different driving intensities. For each intensity, the wavelength dependent yield has been normalized to allow for direct comparison between high and low intensities, whose true yields differ by orders of magnitude. The white lines in the figure indicate the predicted H9 photon energies where direct resonant enhancement would occur according to Eq. (4). Two enhancement features can be recognized in Fig.~\ref{fig:EnhancementMap}, due to the $3p$ and  $2p$ states.  The $2p$ enhancement in general follows Eq. (4), but can be seen to shift less than ponderomotively for higher intensities. The $3p$ enhancement feature follows the prediction of Eq. (4) only for  intensities above approximately $60$ TW/cm$^2$. At lower intensities, the $3p$ feature splits into two, with the lower energy branch marked by the dashed black line. 

The split $3p$ enhancement feature can be understood as a generalized Autler-Townes splitting due to a near-resonant coupling between the $3p$ and $2s$ states induced by the driving laser field. In this case, the dressed states of the laser-driven atom constitutes an Autler-Townes doublet which is approximately symmetric around the $3p$ state  \cite{Cohen-1996}. We note that the separation of the two "states" appearing in the $3p$ enhancement feature is not exactly equal to the Autler-Townes energy separation one would see in a fixed-pump, scanning-probe scenario. This is because in our calculation the pump (the driving laser) and the probe (the 9th harmonic) are locked to each other and thus both vary, which means that each of the dressed states in the enhancement map is 9-photon resonant with the ground state at a different pump wavelength. We believe this is the first time an Autler-Townes splitting has been characterized  using only the harmonic spectrum. 

To summarize this section: we have examined resonant enhancements in the harmonic spectrum using several different representations. We have shown that we can consistently identify and follow resonant enhancements in several harmonics due to the (Stark-shifted) $2p-4p$ states. We also showed that the $3p$ enhancement feature splits into an Autler-Townes like doublet when the driving laser wavelength is such that the $3p$ is near-resonant with the $2s$ state. 


\section{Single Atom: Time Domain} 

\begin{figure}[t!]
		\includegraphics[width=7.5cm,trim = 7mm 0mm 0mm 0mm,clip]{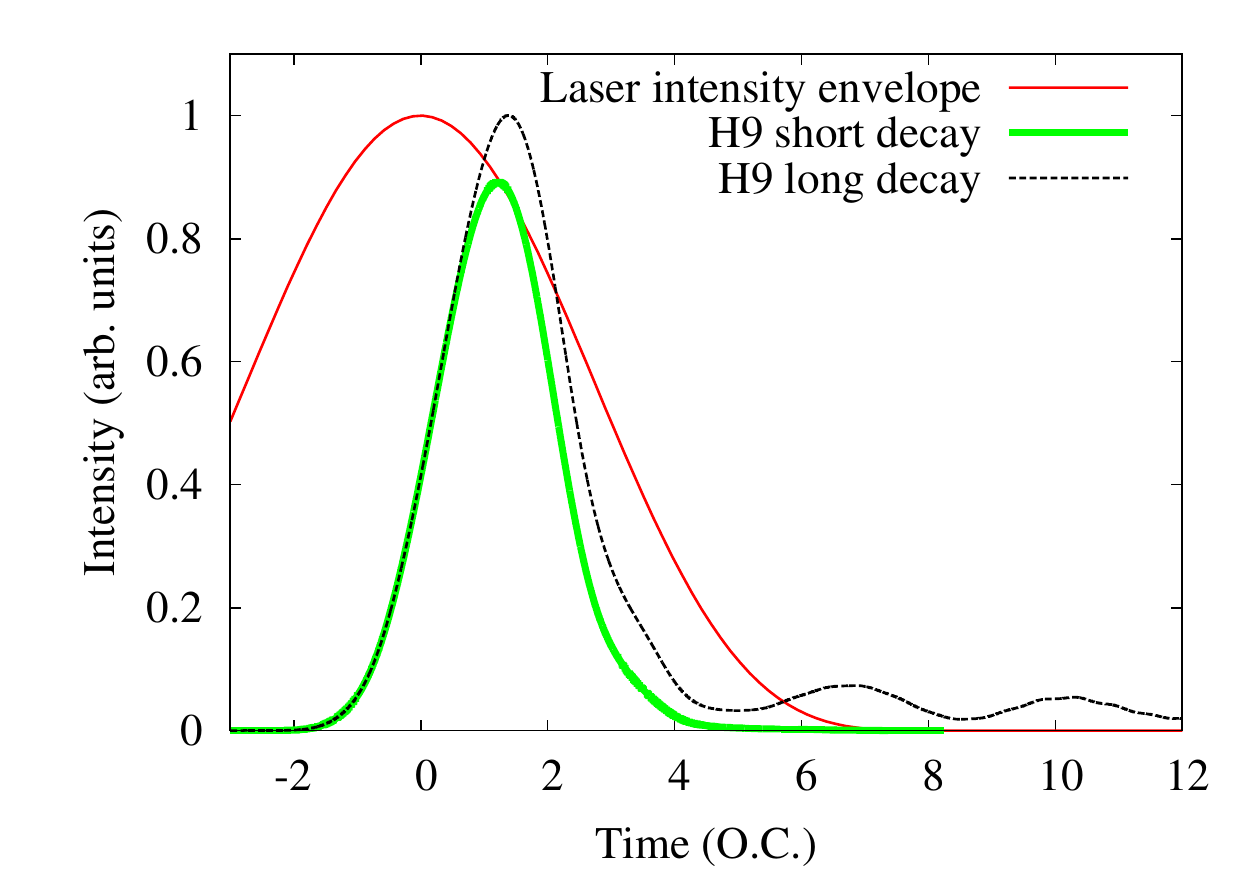}
\caption{H9 time profile at $440$ nm and $140$ TW/cm$^2$ with the (scaled) intensity envelope of the driving pulse, for the short and long Hanning windows.}	
	\label{fig:H9_driven}			
\end{figure}

In this section we concentrate on the time profile of the harmonic emission in the vicinity of resonant enhancement. We will study both the overall envelope of the harmonic pulse, and its sub-cycle time structure. We calculate the time profile by selecting a range of frequencies from the harmonic spectrum and inverse Fourier transforming to the time domain \cite{Scrinzi-2003, Tate-2007}. We start by showing the time-profile of the envelope of H9, calculated by applying a narrow (2$\omega$ wide) frequency filter. Fig. \ref{fig:H9_driven} compares the time profile of $3p$-enhanced H9 at $440$ nm and $140$ TW/cm$^2$ using the short and long Hanning windows, respectively. This comparison shows that the short Hanning window makes very little difference to the driven part of the harmonic response, but that it effectively cuts off the population-driven dipole response, in agreement with what we found in the spectral domain in Fig.~\ref{fig:ringing-spectra}. We note that the oscillatory behavior in the tail of the H9 time profile is due to quantum beating between population left in the $3p$ and $4p$ states which due to their energy difference of 0.7 eV gives rise to a beat period of 6 fs. Quantum beating as a result of a broadband XUV excitation has recently been experimentally observed in neon in a transient absorption scenario \cite{Leone-2014}. 

\begin{figure}[t!]
		\includegraphics[width=7cm]{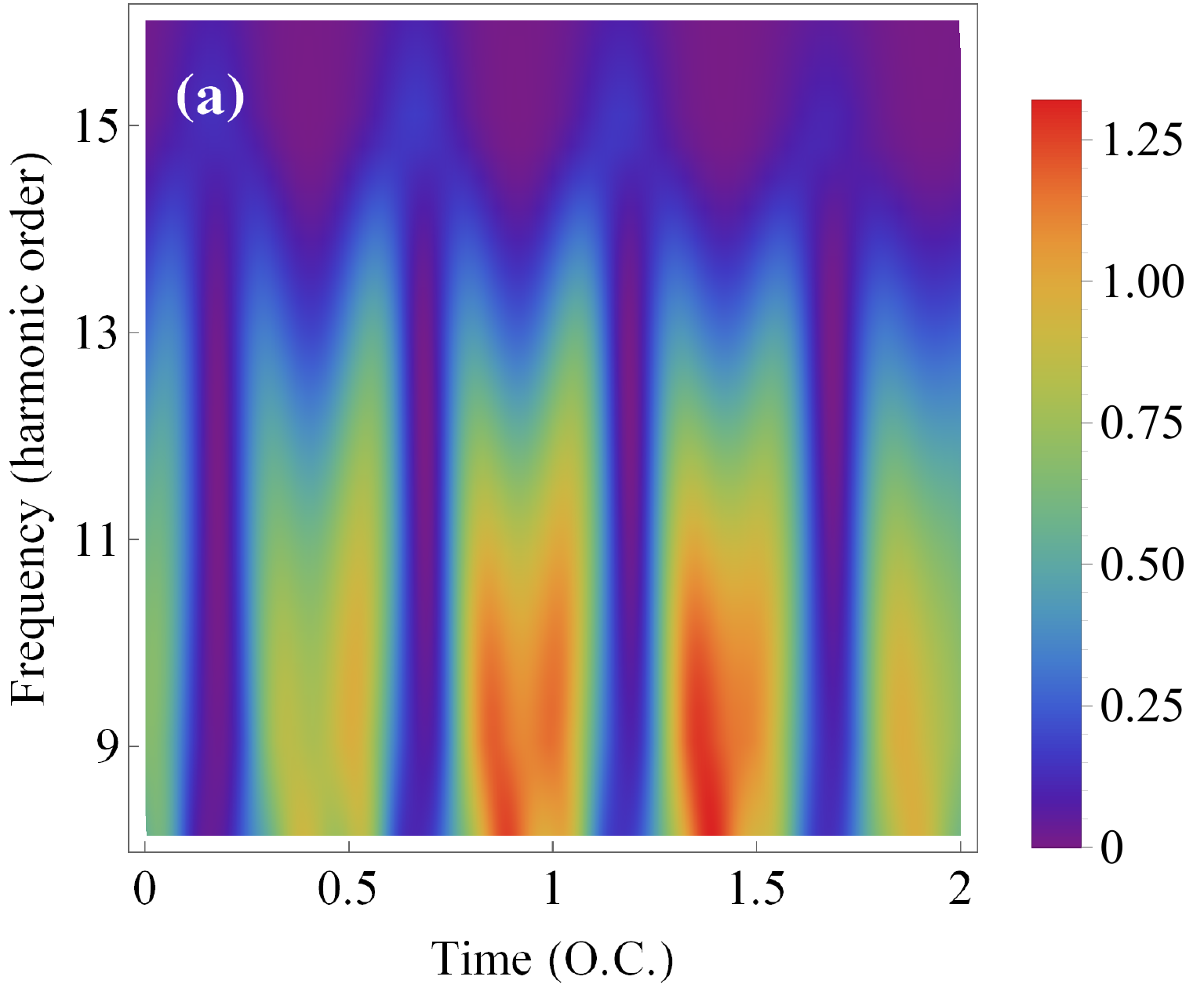}
		\includegraphics[width=7 cm,trim = 8mm 0mm 0mm 0mm,clip]{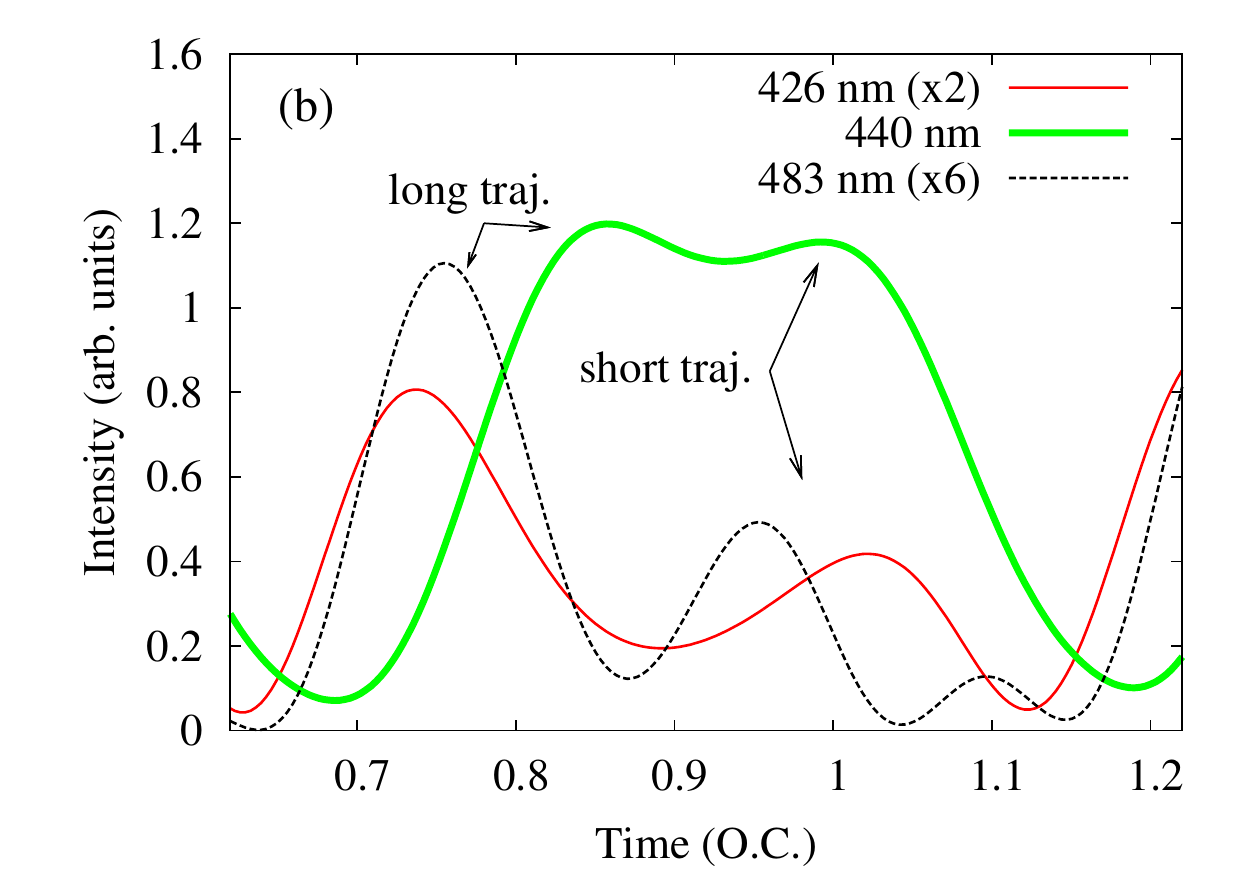}
		\caption{(a) Time-frequency profile of the harmonic radiation for a driving wavelength of 440 nm and  intensity of $140$ TW/cm$^2$, when H9 is resonant with the Stark-shifted $3p$ state. (b) Line-outs of the time-profile of H9 for driving wavelengths of 426 nm, 440 nm, and 483 nm. Note that the long and short trajectory emission peaks in (b) belong to successive half-cycles of the driving field as can be seen in (a).}
	\label{fig:time_freq_plot}
\end{figure}

We next consider the sub-cycle time profile of the harmonic radiation by using a wide ($12\omega$) spectral window. By sliding the central frequency of the window function through the harmonic spectrum, we construct the time-frequency profile of the harmonic radiation. An example of such a time-frequency profile is shown in Fig.~\ref{fig:time_freq_plot}(a) for the on-resonance case of driving wavelength and intensity of 440 nm and 140~TW/cm$^2$. 
The sub-cycle time-frequency profile clearly shows that the harmonic generation process,  even for these low-order harmonics, is dominated by the well-known three-step model dynamics which leads to two dominant emission peaks per half-cycle, with the peaks getting closer in time for higher order harmonics. This is in agreement with previous findings \cite{Yost-2009, Tate-2010, Soifer-2010, Botheron-2010, Li-2014}. Note that although the field-free ionization threshold for this wavelength is just below H9, the Stark-shifted ionization threshold at the peak of the pulse is well above H9, and the 9th harmonic is thus a below-threshold harmonic for most of the duration of the pulse. 

Fig.~\ref{fig:time_freq_plot}(b) shows line-outs centered on H9 in the time-frequency plot, for three different driving wavelengths (scaled for better comparison). The long and short trajectory peaks (marked in the figure) have been identified from time-frequency profiles such as that shown in Fig.~\ref{fig:time_freq_plot}(a). For the short trajectory contribution, the figure shows that as the wavelength increases the emission peak shifts to earlier in the cycle. This is as one would expect from the positive atto-chirp of the short trajectory contribution \cite{Doumy-2009}. The negative atto-chirp of the long-trajectory contribution predicts that the peak would shift later in the cycle as the wavelength increases, which is true for the two non-resonant wavelengths. However, on resonance the long trajectory peak is shifted sharply later in the cycle by approximately 1/8 O.C, corresponding to a phase shift of $\pi/4$. 

\begin{figure}[t!]
	\centering
	 \scalebox{.5}{
		\includegraphics{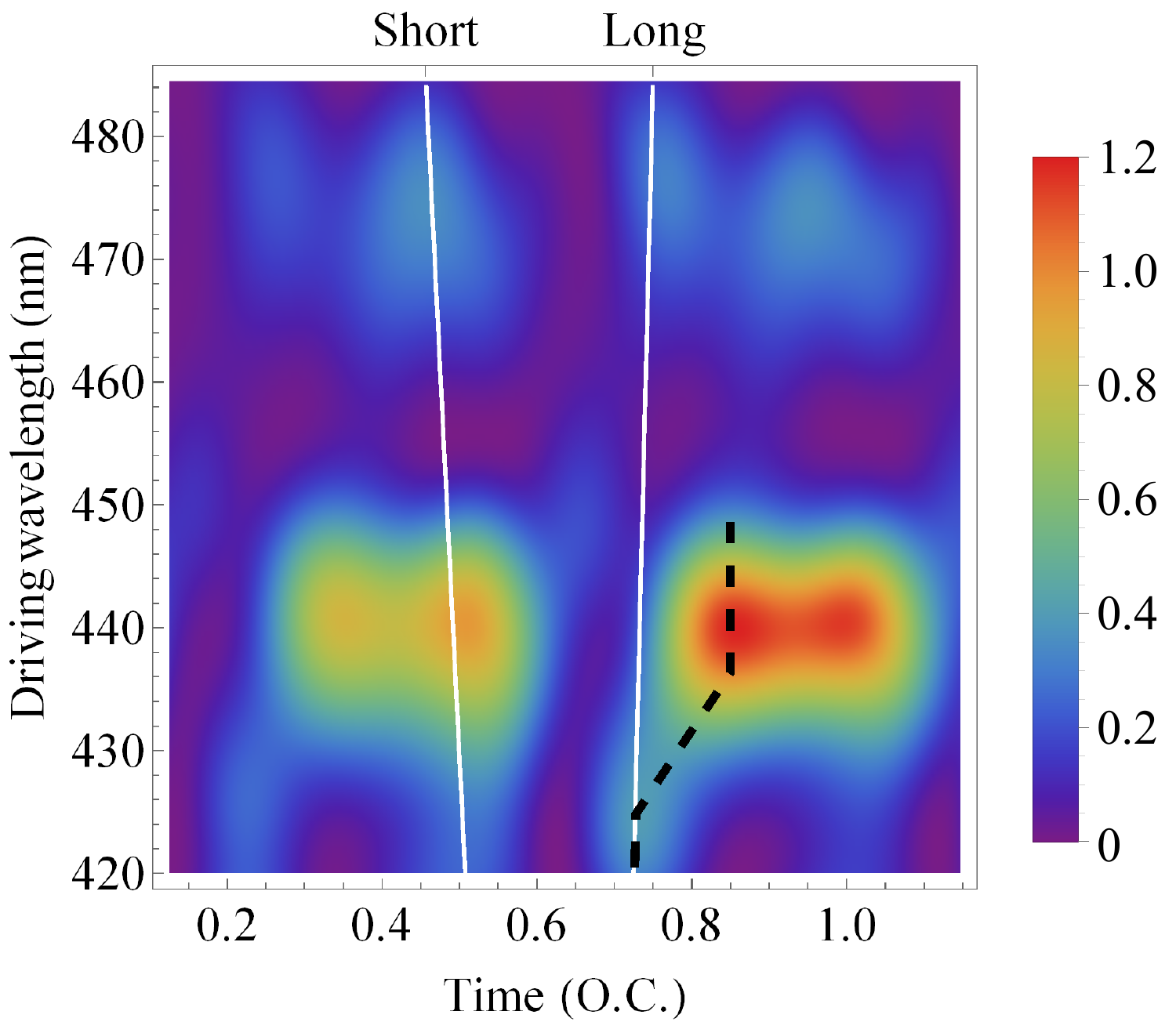}}
		\caption{Sub-cycle time profile of the radiation centered on H$9$ as a function of  driving wavelengths spanning the $3p$ and $2p$ resonant enhancement features, which are located at 440 nm and 470 nm for this intensity. The white lines are an approximate measure of the expected emission times for the short and long trajectory emission peaks. }
	\label{fig:Chirp_Plot}
\end{figure}

We can follow the phase shift of the long trajectory peak near resonance by plotting the H9 sub-cycle time profile as a function of driving wavelength, as shown in Fig.~\ref{fig:Chirp_Plot}. The short and long trajectory peaks have again been identified from time-frequency profiles at different driving wavelengths. The white lines have been added as an approximate measure of the atto-chirp, by connecting the emission times at the shortest and longest wavelengths that are both non-resonant (note that the traditional atto-chirp is not defined for the harmonics below the field-free threshold). As the wavelength increases from approximately 425 nm to 440 nm, the long trajectory emission peak shifts later in time by approximately 1/8 O.C. Between the $3p$ and $2p$ resonances at 440 nm and 475 nm, respectively, the harmonic yield is weak and there are more than two emissions peaks per half O.C., which makes it difficult to identify a short and long trajectory contribution. However, as the wavelength increases toward the $2p$ resonance at 475 nm, the short and long trajectory emission peaks are again clearly identifiable. Although we cannot follow the quantum paths in the wavelength region below the $2p$ resonance at 475 nm, the shift of the long trajectory peak appears to be smaller than for the $3p$ resonance. 

We find similar results for H11 at longer wavelengths (not shown in the figure), when it is in resonance with the $2p-5p$ states. As the (Stark-shifted) $4p$ and $5p$ states come into resonance around 490 nm, the long trajectory emission peak is shifted later in time by approximately 1/10 O.C and stays shifted through the $3p$ resonant enhancement at 505 nm. For wavelengths between the $3p$ and $2p$ resonances (515-530 nm), the time profile is dominated by emission peaks that are not easily identifiable as long or short trajectory contributions, but as H11 comes into resonance with the Stark-shifted $2p$ state around 540 nm, these peaks reappear clearly in the time profile. 

The results in Figs.~\ref{fig:time_freq_plot} and \ref{fig:Chirp_Plot} allow us to draw a number of conclusions: (i) Harmonics that are enhanced by bound state resonances still exhibit strong features that correlate with the well-known short and long re-colliding electron trajectories. Our results in fact suggest that the short and long trajectory contributions are more easily identified close to resonance than in between. (ii) Both quantum path contributions are enhanced around the resonances, indicating that the enhancement happens predominantly  in the ionization stage, via a multiphoton resonance between the driving field and a Stark-shifted resonance. (iii) The fact that only the long trajectory emission peak is shifted in time suggests that this phase shift is imposed in the recollision stage of the harmonic generation process - whereas the short trajectory emission happens very shortly after the ionization time, the long trajectory wave packet spends a long time in the continuum and samples the full spatial range of the ionic potential upon return. Our results also suggest that this phase shift is specific to a particular resonance (the $3p$ state), since both H9 and H11 show exhibit similar phase shifts around the $3p$ resonance and almost no phase shift around the $2p$ resonance. 

To conclude this section on the single-atom emission times, it is interesting to note that on resonance, the envelope of the H9 emission is shifted slightly later in time than off resonance. This can be seen both in Fig.~\ref{fig:time_freq_plot}(a), where the most intense pair of short/long emission peaks is centered around 1.2 O.C, and in Fig.~\ref{fig:H9_driven} as the position of the envelope of the H9-only radiation which also peaks around 1.2 O.C. Off resonance, the envelope peaks closer to 0.7 O.C. (not shown in the figure).  We also see this shift of the envelope for H11 around the $np$ resonances. It is possible that this overall delay could be due to trapping of the electron in the excited state prior to ionization (delayed ionization), so that the excited electrons are still launched into the continuum at the times that lead to return along the short or long trajectories, respectively, but half a cycle later. Such a scenario was discussed in \cite{Mauger-2012} in the context of double ionization, in which an excited electron is caught in a so-called sticky region characterized by resonances in the combined laser-ion potential. We note that it is {\it not} likely that the overall delay is caused by the electron being trapped in the excited state upon returning to the vicinity of the core - this would generally give rise to emission times that would not correspond to those of the short and long trajectory return times.

\section{Macroscopic response}

Finally, we consider the macroscopic harmonic response, in particular whether and how the resonant enhancement manifests itself in this respect. The open triangles in Fig. \ref{fig:SA_vs_Macro} show the macroscopic yield of H9 as a function of wavelength, in direct comparison with the single-atom result (solid diamonds). The shape of the macroscopic curve agrees well with the shape of the single atom curve, the macroscopic results reproducing the strong $3p$ enhancement feature and the weaker $2p$ enhancement feature, however, the macroscopic enhancement features are shifted slightly to longer wavelengths. This difference between the single atom and macroscopic calculations is a consequence of the focal volume averaging of the intensity that is present in the macroscopic calculation.  Looking at Fig. \ref{fig:EnhancementMap}, we see that at lower driving intensity, the resonant wavelength for both the $3p$ and $2p$ single-atom enhancement features shifts towards longer wavelengths due to the change in the Stark shift. Overall, we conclude from Fig. \ref{fig:SA_vs_Macro} that the enhancements  found at the single atom level do indeed survive the macroscopic propagation. 

\begin{figure}[t]
		\includegraphics[width=7.5cm,trim = 8mm 0mm 0mm 0mm,clip]{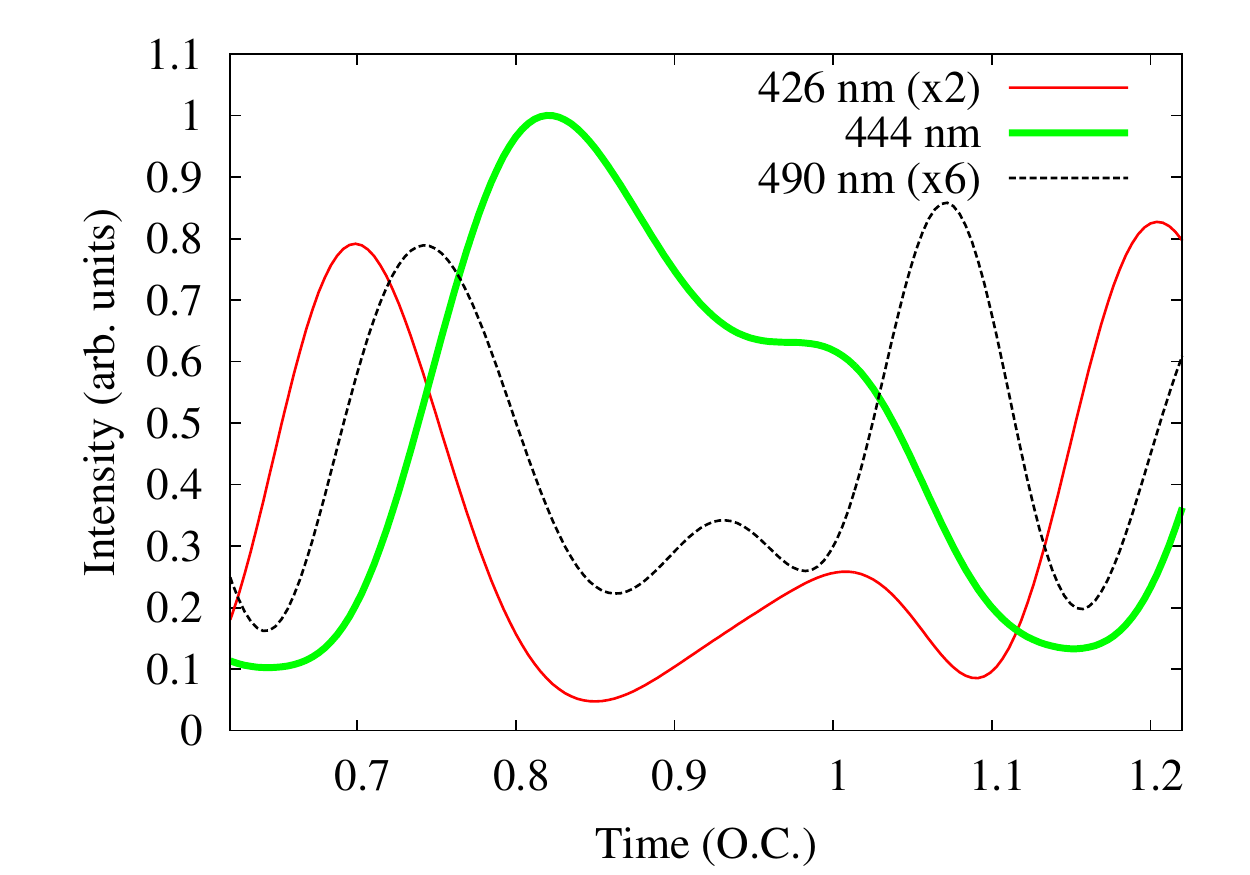}
		\caption{Sub-cycle  time profiles for the macroscopic H$9$ yield, for three different driving wavelengths.}
	\label{fig:Macro-Atto_WavelengthCompare}
\end{figure}

In Fig. \ref{fig:Macro-Atto_WavelengthCompare}, we compare the macroscopic sub-cycle time profiles centered on H9 for three different driving wavelengths. We have slightly shifted the driving wavelengths from the single atom results of Fig. \ref{fig:time_freq_plot}(b) from $440$ nm to $444$ nm and $483$ nm to $490$ nm in order to compensate for the shift of the enhancement peaks to longer wavelengths in our macroscopic calculation as discussed above. We find good agreement between Fig. \ref{fig:Macro-Atto_WavelengthCompare} and Fig. \ref{fig:time_freq_plot}(b). While there are some distortions in the macroscopic case, the long and short trajectory peaks for all driving wavelengths of Fig. \ref{fig:time_freq_plot}(b) are recognizable in Fig. \ref{fig:Macro-Atto_WavelengthCompare} at the corresponding shifted wavelength. The phase shift of the long trajectory peak is also still clearly recognizable in the macroscopic response. One difference between the single atom  and the macroscopic results is the relative enhancement of the long and short trajectory contributions. In the single atom case, both quantum path contributions are equally enhanced, whereas in the macroscopic case, the long trajectory contribution is relatively stronger, by approximately a factor of two. This can be understood by looking at the single atom calculations in Fig.~\ref{fig:Chirp_Plot}. For intensities slightly lower than the resonant peak intensity, the long trajectory contribution is already both enhanced and shifted compared to the short trajectory contribution. Since the macroscopic signal results from the radially integrated harmonic yield (i.e. an integration over lower-intensity contributions), this favors the long trajectory contribution. Note that we have tested that this effect is not dependent on phase matching. The relative position of the laser focus and the center of the gas jet was chosen so as to optimize phase matching of the short trajectory, but in this case we in fact find little difference in the macroscopic yield or time profiles when we move the focus to the center of the jet.


\section{Summary}

We have presented a study of the interplay between resonant enhancement and quantum path dynamics in near-threshold harmonic generation in helium. We concentrated on the driven harmonic-generation response by time-filtering the harmonic signal so as to suppress the long-lasting radiation that would result from population left in excited states at the end of the pulse. By varying the wavelength and intensity of the near-visible driving laser field, we have identified a number of direct and indirect enhancements of H7, H9, and H11 via the Stark-shifted $2p-5p$ states. For H9, we observed an Autler-Townes-like splitting of the enhancement feature due to the $3p$ state, when the wavelength and intensity are such that the driving field strongly couples the $3p$ state to the nearby dark $2s$ state. 

In terms of the quantum path dynamics, we found that both the short and long trajectory contributions to the harmonic emission can be easily identified for harmonics that are resonantly enhanced via the Stark-shifted $np$ states. We found that both contributions are enhanced on resonance, and that the maximum of the envelope of the resonant harmonic is delayed by approximately 0.5 O.C. We interpret this to mean that the enhancement happens via a multiphoton resonance between the ground state and the Stark-shifted excited state, and that the electron is then trapped for a while in the excited state before entering the continuum. Furthermore, we found that only the long trajectory contribution  acquires a phase shift, which leads to a delay in emission time of approximately 0.125 O.C, suggesting that the phase shift is acquired in the interaction between the returning electron wave packet and the ion core for which there is a large difference in the short and long trajectory dynamics. Finally, we showed that both the enhancement and the phase shift are still visible in the macroscopic response. This means that these effects could potentially be explored experimentally, especially considering that our  calculations predict that the macroscopic response is dominated by the long trajectory contribution which exhibits the on-resonance phase shift.



\acknowledgements

We acknowledge discussions with F. Mauger about delayed ionization. This work was supported by the National Science Foundation under Grant No. PHY-1403236. Portions of this research were conducted with high performance computing resources provided by Louisiana State University.  

\bibliography{REH-Paper2}

\begin{thebibliography}{41}%
\makeatletter
\providecommand \@ifxundefined [1]{%
 \@ifx{#1\undefined}
}%
\providecommand \@ifnum [1]{%
 \ifnum #1\expandafter \@firstoftwo
 \else \expandafter \@secondoftwo
 \fi
}%
\providecommand \@ifx [1]{%
 \ifx #1\expandafter \@firstoftwo
 \else \expandafter \@secondoftwo
 \fi
}%
\providecommand \natexlab [1]{#1}%
\providecommand \enquote  [1]{``#1''}%
\providecommand \bibnamefont  [1]{#1}%
\providecommand \bibfnamefont [1]{#1}%
\providecommand \citenamefont [1]{#1}%
\providecommand \href@noop [0]{\@secondoftwo}%
\providecommand \href [0]{\begingroup \@sanitize@url \@href}%
\providecommand \@href[1]{\@@startlink{#1}\@@href}%
\providecommand \@@href[1]{\endgroup#1\@@endlink}%
\providecommand \@sanitize@url [0]{\catcode `\\12\catcode `\$12\catcode
  `\&12\catcode `\#12\catcode `\^12\catcode `\_12\catcode `\%12\relax}%
\providecommand \@@startlink[1]{}%
\providecommand \@@endlink[0]{}%
\providecommand \url  [0]{\begingroup\@sanitize@url \@url }%
\providecommand \@url [1]{\endgroup\@href {#1}{\urlprefix }}%
\providecommand \urlprefix  [0]{URL }%
\providecommand \Eprint [0]{\href }%
\providecommand \doibase [0]{http://dx.doi.org/}%
\providecommand \selectlanguage [0]{\@gobble}%
\providecommand \bibinfo  [0]{\@secondoftwo}%
\providecommand \bibfield  [0]{\@secondoftwo}%
\providecommand \translation [1]{[#1]}%
\providecommand \BibitemOpen [0]{}%
\providecommand \bibitemStop [0]{}%
\providecommand \bibitemNoStop [0]{.\EOS\space}%
\providecommand \EOS [0]{\spacefactor3000\relax}%
\providecommand \BibitemShut  [1]{\csname bibitem#1\endcsname}%
\let\auto@bib@innerbib\@empty
\bibitem [{\citenamefont {McPherson}\ \emph {et~al.}(1987)\citenamefont
  {McPherson}, \citenamefont {Gibson}, \citenamefont {Jara}, \citenamefont
  {Johann}, \citenamefont {Luk}, \citenamefont {McIntyre}, \citenamefont
  {Boyer},\ and\ \citenamefont {Rhodes}}]{McPherson-1987}%
  \BibitemOpen
  \bibfield  {author} {\bibinfo {author} {\bibfnamefont {A.}~\bibnamefont
  {McPherson}}, \bibinfo {author} {\bibfnamefont {G.}~\bibnamefont {Gibson}},
  \bibinfo {author} {\bibfnamefont {H.}~\bibnamefont {Jara}}, \bibinfo {author}
  {\bibfnamefont {U.}~\bibnamefont {Johann}}, \bibinfo {author} {\bibfnamefont
  {T.~S.}\ \bibnamefont {Luk}}, \bibinfo {author} {\bibfnamefont {I.~A.}\
  \bibnamefont {McIntyre}}, \bibinfo {author} {\bibfnamefont {K.}~\bibnamefont
  {Boyer}}, \ and\ \bibinfo {author} {\bibfnamefont {C.~K.}\ \bibnamefont
  {Rhodes}},\ }\href {\doibase 10.1364/JOSAB.4.000595} {\bibfield  {journal}
  {\bibinfo  {journal} {J. Opt. Soc. Am. B}\ }\textbf {\bibinfo {volume} {4}},\
  \bibinfo {pages} {595} (\bibinfo {year} {1987})}\BibitemShut {NoStop}%
\bibitem [{\citenamefont {Ferray}\ \emph {et~al.}(1988)\citenamefont {Ferray},
  \citenamefont {L'Huillier}, \citenamefont {Li}, \citenamefont {Lompre},
  \citenamefont {Mainfray},\ and\ \citenamefont {Manus}}]{Ferray-1988}%
  \BibitemOpen
  \bibfield  {author} {\bibinfo {author} {\bibfnamefont {M.}~\bibnamefont
  {Ferray}}, \bibinfo {author} {\bibfnamefont {A.}~\bibnamefont {L'Huillier}},
  \bibinfo {author} {\bibfnamefont {X.~F.}\ \bibnamefont {Li}}, \bibinfo
  {author} {\bibfnamefont {L.~A.}\ \bibnamefont {Lompre}}, \bibinfo {author}
  {\bibfnamefont {G.}~\bibnamefont {Mainfray}}, \ and\ \bibinfo {author}
  {\bibfnamefont {C.}~\bibnamefont {Manus}},\ }\href
  {http://stacks.iop.org/0953-4075/21/i=3/a=001} {\bibfield  {journal}
  {\bibinfo  {journal} {Journal of Physics B: Atomic, Molecular and Optical
  Physics}\ }\textbf {\bibinfo {volume} {21}},\ \bibinfo {pages} {L31}
  (\bibinfo {year} {1988})}\BibitemShut {NoStop}%
\bibitem [{\citenamefont {Schafer}\ \emph {et~al.}(1993)\citenamefont
  {Schafer}, \citenamefont {Yang}, \citenamefont {DiMauro},\ and\ \citenamefont
  {Kulander}}]{Schafer-1993}%
  \BibitemOpen
  \bibfield  {author} {\bibinfo {author} {\bibfnamefont {K.~J.}\ \bibnamefont
  {Schafer}}, \bibinfo {author} {\bibfnamefont {B.}~\bibnamefont {Yang}},
  \bibinfo {author} {\bibfnamefont {L.~F.}\ \bibnamefont {DiMauro}}, \ and\
  \bibinfo {author} {\bibfnamefont {K.~C.}\ \bibnamefont {Kulander}},\ }\href
  {\doibase 10.1103/PhysRevLett.70.1599} {\bibfield  {journal} {\bibinfo
  {journal} {Phys. Rev. Lett.}\ }\textbf {\bibinfo {volume} {70}},\ \bibinfo
  {pages} {1599} (\bibinfo {year} {1993})}\BibitemShut {NoStop}%
\bibitem [{\citenamefont {Corkum}(1993)}]{Corkum-1994}%
  \BibitemOpen
  \bibfield  {author} {\bibinfo {author} {\bibfnamefont {P.~B.}\ \bibnamefont
  {Corkum}},\ }\href {\doibase 10.1103/PhysRevLett.71.1994} {\bibfield
  {journal} {\bibinfo  {journal} {Phys. Rev. Lett.}\ }\textbf {\bibinfo
  {volume} {71}},\ \bibinfo {pages} {1994} (\bibinfo {year}
  {1993})}\BibitemShut {NoStop}%
\bibitem [{\citenamefont {Itatani}\ \emph {et~al.}(2004)\citenamefont
  {Itatani}, \citenamefont {Levesque}, \citenamefont {Zeidler}, \citenamefont
  {Niikura}, \citenamefont {P{\'e}pin}, \citenamefont {Kieffer}, \citenamefont
  {Corkum},\ and\ \citenamefont {Villeneuve}}]{Itatani-2004}%
  \BibitemOpen
  \bibfield  {author} {\bibinfo {author} {\bibfnamefont {J.}~\bibnamefont
  {Itatani}}, \bibinfo {author} {\bibfnamefont {J.}~\bibnamefont {Levesque}},
  \bibinfo {author} {\bibfnamefont {D.}~\bibnamefont {Zeidler}}, \bibinfo
  {author} {\bibfnamefont {H.}~\bibnamefont {Niikura}}, \bibinfo {author}
  {\bibfnamefont {H.}~\bibnamefont {P{\'e}pin}}, \bibinfo {author}
  {\bibfnamefont {J.-C.}\ \bibnamefont {Kieffer}}, \bibinfo {author}
  {\bibfnamefont {P.~B.}\ \bibnamefont {Corkum}}, \ and\ \bibinfo {author}
  {\bibfnamefont {D.~M.}\ \bibnamefont {Villeneuve}},\ }\href@noop {}
  {\bibfield  {journal} {\bibinfo  {journal} {Nature}\ }\textbf {\bibinfo
  {volume} {432}},\ \bibinfo {pages} {867} (\bibinfo {year}
  {2004})}\BibitemShut {NoStop}%
\bibitem [{\citenamefont {Morishita}\ \emph {et~al.}(2008)\citenamefont
  {Morishita}, \citenamefont {Le}, \citenamefont {Chen},\ and\ \citenamefont
  {Lin}}]{Morishita-2008}%
  \BibitemOpen
  \bibfield  {author} {\bibinfo {author} {\bibfnamefont {T.}~\bibnamefont
  {Morishita}}, \bibinfo {author} {\bibfnamefont {A.-T.}\ \bibnamefont {Le}},
  \bibinfo {author} {\bibfnamefont {Z.}~\bibnamefont {Chen}}, \ and\ \bibinfo
  {author} {\bibfnamefont {C.~D.}\ \bibnamefont {Lin}},\ }\href {\doibase
  10.1103/PhysRevLett.100.013903} {\bibfield  {journal} {\bibinfo  {journal}
  {Phys. Rev. Lett.}\ }\textbf {\bibinfo {volume} {100}},\ \bibinfo {pages}
  {013903} (\bibinfo {year} {2008})}\BibitemShut {NoStop}%
\bibitem [{\citenamefont {Le}\ \emph {et~al.}(2008)\citenamefont {Le},
  \citenamefont {Morishita},\ and\ \citenamefont {Lin}}]{Le-2008}%
  \BibitemOpen
  \bibfield  {author} {\bibinfo {author} {\bibfnamefont {A.-T.}\ \bibnamefont
  {Le}}, \bibinfo {author} {\bibfnamefont {T.}~\bibnamefont {Morishita}}, \
  and\ \bibinfo {author} {\bibfnamefont {C.~D.}\ \bibnamefont {Lin}},\ }\href
  {\doibase 10.1103/PhysRevA.78.023814} {\bibfield  {journal} {\bibinfo
  {journal} {Phys. Rev. A}\ }\textbf {\bibinfo {volume} {78}},\ \bibinfo
  {pages} {023814} (\bibinfo {year} {2008})}\BibitemShut {NoStop}%
\bibitem [{\citenamefont {Smirnova}\ \emph {et~al.}(2009)\citenamefont
  {Smirnova}, \citenamefont {Mairesse}, \citenamefont {Patchkovskii},
  \citenamefont {Dudovich}, \citenamefont {Villeneuve}, \citenamefont
  {Corkum},\ and\ \citenamefont {Ivanov}}]{Smirnova-2009}%
  \BibitemOpen
  \bibfield  {author} {\bibinfo {author} {\bibfnamefont {O.}~\bibnamefont
  {Smirnova}}, \bibinfo {author} {\bibfnamefont {Y.}~\bibnamefont {Mairesse}},
  \bibinfo {author} {\bibfnamefont {S.}~\bibnamefont {Patchkovskii}}, \bibinfo
  {author} {\bibfnamefont {N.}~\bibnamefont {Dudovich}}, \bibinfo {author}
  {\bibfnamefont {D.}~\bibnamefont {Villeneuve}}, \bibinfo {author}
  {\bibfnamefont {P.}~\bibnamefont {Corkum}}, \ and\ \bibinfo {author}
  {\bibfnamefont {M.~Y.}\ \bibnamefont {Ivanov}},\ }\href@noop {} {\bibfield
  {journal} {\bibinfo  {journal} {Nature}\ }\textbf {\bibinfo {volume} {460}},\
  \bibinfo {pages} {972} (\bibinfo {year} {2009})}\BibitemShut {NoStop}%
\bibitem [{\citenamefont {W{\"o}rner}\ \emph {et~al.}(2010)\citenamefont
  {W{\"o}rner}, \citenamefont {Bertrand}, \citenamefont {Kartashov},
  \citenamefont {Corkum},\ and\ \citenamefont {Villeneuve}}]{Worner-2010}%
  \BibitemOpen
  \bibfield  {author} {\bibinfo {author} {\bibfnamefont {H.}~\bibnamefont
  {W{\"o}rner}}, \bibinfo {author} {\bibfnamefont {J.}~\bibnamefont
  {Bertrand}}, \bibinfo {author} {\bibfnamefont {D.}~\bibnamefont {Kartashov}},
  \bibinfo {author} {\bibfnamefont {P.}~\bibnamefont {Corkum}}, \ and\ \bibinfo
  {author} {\bibfnamefont {D.}~\bibnamefont {Villeneuve}},\ }\href@noop {}
  {\bibfield  {journal} {\bibinfo  {journal} {Nature}\ }\textbf {\bibinfo
  {volume} {466}},\ \bibinfo {pages} {604} (\bibinfo {year}
  {2010})}\BibitemShut {NoStop}%
\bibitem [{\citenamefont {Toma}\ \emph {et~al.}(1999)\citenamefont {Toma},
  \citenamefont {Antoine}, \citenamefont {de~Bohan},\ and\ \citenamefont
  {Muller}}]{Toma-1999}%
  \BibitemOpen
  \bibfield  {author} {\bibinfo {author} {\bibfnamefont {E.~S.}\ \bibnamefont
  {Toma}}, \bibinfo {author} {\bibfnamefont {P.}~\bibnamefont {Antoine}},
  \bibinfo {author} {\bibfnamefont {A.}~\bibnamefont {de~Bohan}}, \ and\
  \bibinfo {author} {\bibfnamefont {H.~G.}\ \bibnamefont {Muller}},\ }\href
  {http://stacks.iop.org/0953-4075/32/i=24/a=318} {\bibfield  {journal}
  {\bibinfo  {journal} {Journal of Physics B: Atomic, Molecular and Optical
  Physics}\ }\textbf {\bibinfo {volume} {32}},\ \bibinfo {pages} {5843}
  (\bibinfo {year} {1999})}\BibitemShut {NoStop}%
\bibitem [{\citenamefont {Gaarde}\ and\ \citenamefont
  {Schafer}(2001)}]{Schafer-2001}%
  \BibitemOpen
  \bibfield  {author} {\bibinfo {author} {\bibfnamefont {M.~B.}\ \bibnamefont
  {Gaarde}}\ and\ \bibinfo {author} {\bibfnamefont {K.~J.}\ \bibnamefont
  {Schafer}},\ }\href {\doibase 10.1103/PhysRevA.64.013820} {\bibfield
  {journal} {\bibinfo  {journal} {Phys. Rev. A}\ }\textbf {\bibinfo {volume}
  {64}},\ \bibinfo {pages} {013820} (\bibinfo {year} {2001})}\BibitemShut
  {NoStop}%
\bibitem [{\citenamefont {Figueira~de Morisson~Faria}\ \emph
  {et~al.}(2002)\citenamefont {Figueira~de Morisson~Faria}, \citenamefont
  {Kopold}, \citenamefont {Becker},\ and\ \citenamefont
  {Rost}}]{Figueira-2002}%
  \BibitemOpen
  \bibfield  {author} {\bibinfo {author} {\bibfnamefont {C.}~\bibnamefont
  {Figueira~de Morisson~Faria}}, \bibinfo {author} {\bibfnamefont
  {R.}~\bibnamefont {Kopold}}, \bibinfo {author} {\bibfnamefont
  {W.}~\bibnamefont {Becker}}, \ and\ \bibinfo {author} {\bibfnamefont {J.~M.}\
  \bibnamefont {Rost}},\ }\href {\doibase 10.1103/PhysRevA.65.023404}
  {\bibfield  {journal} {\bibinfo  {journal} {Phys. Rev. A}\ }\textbf {\bibinfo
  {volume} {65}},\ \bibinfo {pages} {023404} (\bibinfo {year}
  {2002})}\BibitemShut {NoStop}%
\bibitem [{\citenamefont {Ta{\"i}eb}\ \emph {et~al.}(2003)\citenamefont
  {Ta{\"i}eb}, \citenamefont {V\'eniard}, \citenamefont {Wassaf},\ and\
  \citenamefont {Maquet}}]{Taieb-2003}%
  \BibitemOpen
  \bibfield  {author} {\bibinfo {author} {\bibfnamefont {R.}~\bibnamefont
  {Ta{\"i}eb}}, \bibinfo {author} {\bibfnamefont {V.}~\bibnamefont
  {V\'eniard}}, \bibinfo {author} {\bibfnamefont {J.}~\bibnamefont {Wassaf}}, \
  and\ \bibinfo {author} {\bibfnamefont {A.}~\bibnamefont {Maquet}},\ }\href
  {\doibase 10.1103/PhysRevA.68.033403} {\bibfield  {journal} {\bibinfo
  {journal} {Phys. Rev. A}\ }\textbf {\bibinfo {volume} {68}},\ \bibinfo
  {pages} {033403} (\bibinfo {year} {2003})}\BibitemShut {NoStop}%
\bibitem [{\citenamefont {Strelkov}(2010)}]{Strelkov-2010}%
  \BibitemOpen
  \bibfield  {author} {\bibinfo {author} {\bibfnamefont {V.}~\bibnamefont
  {Strelkov}},\ }\href {\doibase 10.1103/PhysRevLett.104.123901} {\bibfield
  {journal} {\bibinfo  {journal} {Phys. Rev. Lett.}\ }\textbf {\bibinfo
  {volume} {104}},\ \bibinfo {pages} {123901} (\bibinfo {year}
  {2010})}\BibitemShut {NoStop}%
\bibitem [{\citenamefont {Ackermann}\ \emph {et~al.}(2012)\citenamefont
  {Ackermann}, \citenamefont {M\"{u}nch},\ and\ \citenamefont
  {Halfmann}}]{Ackermann-2013}%
  \BibitemOpen
  \bibfield  {author} {\bibinfo {author} {\bibfnamefont {P.}~\bibnamefont
  {Ackermann}}, \bibinfo {author} {\bibfnamefont {H.}~\bibnamefont
  {M\"{u}nch}}, \ and\ \bibinfo {author} {\bibfnamefont {T.}~\bibnamefont
  {Halfmann}},\ }\href {\doibase 10.1364/OE.20.013824} {\bibfield  {journal}
  {\bibinfo  {journal} {Opt. Express}\ }\textbf {\bibinfo {volume} {20}},\
  \bibinfo {pages} {13824} (\bibinfo {year} {2012})}\BibitemShut {NoStop}%
\bibitem [{\citenamefont {Redkin}\ \emph {et~al.}(2011)\citenamefont {Redkin},
  \citenamefont {Kodirov},\ and\ \citenamefont {Ganeev}}]{Redkin-2011}%
  \BibitemOpen
  \bibfield  {author} {\bibinfo {author} {\bibfnamefont {P.~V.}\ \bibnamefont
  {Redkin}}, \bibinfo {author} {\bibfnamefont {M.~K.}\ \bibnamefont {Kodirov}},
  \ and\ \bibinfo {author} {\bibfnamefont {R.~A.}\ \bibnamefont {Ganeev}},\
  }\href {\doibase 10.1364/JOSAB.28.000165} {\bibfield  {journal} {\bibinfo
  {journal} {J. Opt. Soc. Am. B}\ }\textbf {\bibinfo {volume} {28}},\ \bibinfo
  {pages} {165} (\bibinfo {year} {2011})}\BibitemShut {NoStop}%
\bibitem [{\citenamefont {Shiner}\ \emph {et~al.}(2011)\citenamefont {Shiner},
  \citenamefont {Schmidt}, \citenamefont {Trallero-Herrero}, \citenamefont
  {W{\"o}rner}, \citenamefont {Patchkovskii}, \citenamefont {Corkum},
  \citenamefont {Kieffer}, \citenamefont {L{\'e}gar{\'e}},\ and\ \citenamefont
  {Villeneuve}}]{Shiner-2011}%
  \BibitemOpen
  \bibfield  {author} {\bibinfo {author} {\bibfnamefont {A.}~\bibnamefont
  {Shiner}}, \bibinfo {author} {\bibfnamefont {B.}~\bibnamefont {Schmidt}},
  \bibinfo {author} {\bibfnamefont {C.}~\bibnamefont {Trallero-Herrero}},
  \bibinfo {author} {\bibfnamefont {H.}~\bibnamefont {W{\"o}rner}}, \bibinfo
  {author} {\bibfnamefont {S.}~\bibnamefont {Patchkovskii}}, \bibinfo {author}
  {\bibfnamefont {P.}~\bibnamefont {Corkum}}, \bibinfo {author} {\bibfnamefont
  {J.}~\bibnamefont {Kieffer}}, \bibinfo {author} {\bibfnamefont
  {F.}~\bibnamefont {L{\'e}gar{\'e}}}, \ and\ \bibinfo {author} {\bibfnamefont
  {D.}~\bibnamefont {Villeneuve}},\ }\href@noop {} {\bibfield  {journal}
  {\bibinfo  {journal} {Nature Physics}\ }\textbf {\bibinfo {volume} {7}},\
  \bibinfo {pages} {464} (\bibinfo {year} {2011})}\BibitemShut {NoStop}%
\bibitem [{\citenamefont {Tudorovskaya}\ and\ \citenamefont
  {Lein}(2011)}]{Lein-2011}%
  \BibitemOpen
  \bibfield  {author} {\bibinfo {author} {\bibfnamefont {M.}~\bibnamefont
  {Tudorovskaya}}\ and\ \bibinfo {author} {\bibfnamefont {M.}~\bibnamefont
  {Lein}},\ }\href {\doibase 10.1103/PhysRevA.84.013430} {\bibfield  {journal}
  {\bibinfo  {journal} {Phys. Rev. A}\ }\textbf {\bibinfo {volume} {84}},\
  \bibinfo {pages} {013430} (\bibinfo {year} {2011})}\BibitemShut {NoStop}%
\bibitem [{\citenamefont {Jin}\ \emph {et~al.}(2012)\citenamefont {Jin},
  \citenamefont {Bertrand}, \citenamefont {Lucchese}, \citenamefont {W\"orner},
  \citenamefont {Corkum}, \citenamefont {Villeneuve}, \citenamefont {Le},\ and\
  \citenamefont {Lin}}]{Jin-2012}%
  \BibitemOpen
  \bibfield  {author} {\bibinfo {author} {\bibfnamefont {C.}~\bibnamefont
  {Jin}}, \bibinfo {author} {\bibfnamefont {J.~B.}\ \bibnamefont {Bertrand}},
  \bibinfo {author} {\bibfnamefont {R.~R.}\ \bibnamefont {Lucchese}}, \bibinfo
  {author} {\bibfnamefont {H.~J.}\ \bibnamefont {W\"orner}}, \bibinfo {author}
  {\bibfnamefont {P.~B.}\ \bibnamefont {Corkum}}, \bibinfo {author}
  {\bibfnamefont {D.~M.}\ \bibnamefont {Villeneuve}}, \bibinfo {author}
  {\bibfnamefont {A.-T.}\ \bibnamefont {Le}}, \ and\ \bibinfo {author}
  {\bibfnamefont {C.~D.}\ \bibnamefont {Lin}},\ }\href {\doibase
  10.1103/PhysRevA.85.013405} {\bibfield  {journal} {\bibinfo  {journal} {Phys.
  Rev. A}\ }\textbf {\bibinfo {volume} {85}},\ \bibinfo {pages} {013405}
  (\bibinfo {year} {2012})}\BibitemShut {NoStop}%
\bibitem [{\citenamefont {Chu}\ and\ \citenamefont
  {Groenenboom}(2013)}]{Chu-2013}%
  \BibitemOpen
  \bibfield  {author} {\bibinfo {author} {\bibfnamefont {X.}~\bibnamefont
  {Chu}}\ and\ \bibinfo {author} {\bibfnamefont {G.~C.}\ \bibnamefont
  {Groenenboom}},\ }\href {\doibase 10.1103/PhysRevA.87.013434} {\bibfield
  {journal} {\bibinfo  {journal} {Phys. Rev. A}\ }\textbf {\bibinfo {volume}
  {87}},\ \bibinfo {pages} {013434} (\bibinfo {year} {2013})}\BibitemShut
  {NoStop}%
\bibitem [{\citenamefont {Xiong}\ \emph {et~al.}(2014)\citenamefont {Xiong},
  \citenamefont {Geng}, \citenamefont {Tang}, \citenamefont {Peng},\ and\
  \citenamefont {Gong}}]{Gong-2013}%
  \BibitemOpen
  \bibfield  {author} {\bibinfo {author} {\bibfnamefont {W.-H.}\ \bibnamefont
  {Xiong}}, \bibinfo {author} {\bibfnamefont {J.-W.}\ \bibnamefont {Geng}},
  \bibinfo {author} {\bibfnamefont {J.-Y.}\ \bibnamefont {Tang}}, \bibinfo
  {author} {\bibfnamefont {L.-Y.}\ \bibnamefont {Peng}}, \ and\ \bibinfo
  {author} {\bibfnamefont {Q.}~\bibnamefont {Gong}},\ }\href {\doibase
  10.1103/PhysRevLett.112.233001} {\bibfield  {journal} {\bibinfo  {journal}
  {Phys. Rev. Lett.}\ }\textbf {\bibinfo {volume} {112}},\ \bibinfo {pages}
  {233001} (\bibinfo {year} {2014})}\BibitemShut {NoStop}%
\bibitem [{\citenamefont {Chini}\ \emph {et~al.}(2014)\citenamefont {Chini},
  \citenamefont {Wang}, \citenamefont {Cheng}, \citenamefont {Wang},
  \citenamefont {Wu}, \citenamefont {Cunningham}, \citenamefont {Li},
  \citenamefont {Heslar}, \citenamefont {Telnov}, \citenamefont {Chu} \emph
  {et~al.}}]{Chini-2014}%
  \BibitemOpen
  \bibfield  {author} {\bibinfo {author} {\bibfnamefont {M.}~\bibnamefont
  {Chini}}, \bibinfo {author} {\bibfnamefont {X.}~\bibnamefont {Wang}},
  \bibinfo {author} {\bibfnamefont {Y.}~\bibnamefont {Cheng}}, \bibinfo
  {author} {\bibfnamefont {H.}~\bibnamefont {Wang}}, \bibinfo {author}
  {\bibfnamefont {Y.}~\bibnamefont {Wu}}, \bibinfo {author} {\bibfnamefont
  {E.}~\bibnamefont {Cunningham}}, \bibinfo {author} {\bibfnamefont {P.-C.}\
  \bibnamefont {Li}}, \bibinfo {author} {\bibfnamefont {J.}~\bibnamefont
  {Heslar}}, \bibinfo {author} {\bibfnamefont {D.~A.}\ \bibnamefont {Telnov}},
  \bibinfo {author} {\bibfnamefont {S.-I.}\ \bibnamefont {Chu}},  \emph
  {et~al.},\ }\href@noop {} {\bibfield  {journal} {\bibinfo  {journal} {Nature
  Photonics}\ }\textbf {\bibinfo {volume} {8}},\ \bibinfo {pages} {437}
  (\bibinfo {year} {2014})}\BibitemShut {NoStop}%
\bibitem [{\citenamefont {Ferr{\'e}}\ \emph {et~al.}(2015)\citenamefont
  {Ferr{\'e}}, \citenamefont {Boguslavskiy}, \citenamefont {Dagan},
  \citenamefont {Blanchet}, \citenamefont {Bruner}, \citenamefont {Burgy},
  \citenamefont {Camper}, \citenamefont {Descamps}, \citenamefont {Fabre},
  \citenamefont {Fedorov} \emph {et~al.}}]{Ferre-2014}%
  \BibitemOpen
  \bibfield  {author} {\bibinfo {author} {\bibfnamefont {A.}~\bibnamefont
  {Ferr{\'e}}}, \bibinfo {author} {\bibfnamefont {A.}~\bibnamefont
  {Boguslavskiy}}, \bibinfo {author} {\bibfnamefont {M.}~\bibnamefont {Dagan}},
  \bibinfo {author} {\bibfnamefont {V.}~\bibnamefont {Blanchet}}, \bibinfo
  {author} {\bibfnamefont {B.}~\bibnamefont {Bruner}}, \bibinfo {author}
  {\bibfnamefont {F.}~\bibnamefont {Burgy}}, \bibinfo {author} {\bibfnamefont
  {A.}~\bibnamefont {Camper}}, \bibinfo {author} {\bibfnamefont
  {D.}~\bibnamefont {Descamps}}, \bibinfo {author} {\bibfnamefont
  {B.}~\bibnamefont {Fabre}}, \bibinfo {author} {\bibfnamefont
  {N.}~\bibnamefont {Fedorov}},  \emph {et~al.},\ }\href@noop {} {\bibfield
  {journal} {\bibinfo  {journal} {Nature Communications}\ }\textbf {\bibinfo
  {volume} {6}} (\bibinfo {year} {2015})}\BibitemShut {NoStop}%
\bibitem [{\citenamefont {Li}\ \emph {et~al.}(2014)\citenamefont {Li},
  \citenamefont {Sheu}, \citenamefont {Laughlin},\ and\ \citenamefont
  {Chu}}]{Li-2014}%
  \BibitemOpen
  \bibfield  {author} {\bibinfo {author} {\bibfnamefont {P.-C.}\ \bibnamefont
  {Li}}, \bibinfo {author} {\bibfnamefont {Y.-L.}\ \bibnamefont {Sheu}},
  \bibinfo {author} {\bibfnamefont {C.}~\bibnamefont {Laughlin}}, \ and\
  \bibinfo {author} {\bibfnamefont {S.-I.}\ \bibnamefont {Chu}},\ }\href
  {\doibase 10.1103/PhysRevA.90.041401} {\bibfield  {journal} {\bibinfo
  {journal} {Phys. Rev. A}\ }\textbf {\bibinfo {volume} {90}},\ \bibinfo
  {pages} {041401} (\bibinfo {year} {2014})}\BibitemShut {NoStop}%
\bibitem [{\citenamefont {Schafer}(2008)}]{Schafer-Strongfield}%
  \BibitemOpen
  \bibfield  {author} {\bibinfo {author} {\bibfnamefont {K.~J.}\ \bibnamefont
  {Schafer}},\ }in\ \href@noop {} {\emph {\bibinfo {booktitle} {Strong Field
  Laser Physics}}},\ \bibinfo {editor} {edited by\ \bibinfo {editor}
  {\bibfnamefont {T.}~\bibnamefont {Brabec}}}\ (\bibinfo  {publisher} {Springer
  Series in Optical Science},\ \bibinfo {year} {2008})\BibitemShut {NoStop}%
\bibitem [{\citenamefont {Barth}\ and\ \citenamefont
  {Lasser}(2009)}]{Barth-2009}%
  \BibitemOpen
  \bibfield  {author} {\bibinfo {author} {\bibfnamefont {I.}~\bibnamefont
  {Barth}}\ and\ \bibinfo {author} {\bibfnamefont {C.}~\bibnamefont {Lasser}},\
  }\href@noop {} {\bibfield  {journal} {\bibinfo  {journal} {Journal of Physics
  B: Atomic, Molecular and Optical Physics}\ }\textbf {\bibinfo {volume}
  {42}},\ \bibinfo {pages} {235101} (\bibinfo {year} {2009})}\BibitemShut
  {NoStop}%
\bibitem [{\citenamefont {Blackman}\ and\ \citenamefont
  {Tukey}(1958)}]{Blackman-1958}%
  \BibitemOpen
  \bibfield  {author} {\bibinfo {author} {\bibfnamefont {R.~B.}\ \bibnamefont
  {Blackman}}\ and\ \bibinfo {author} {\bibfnamefont {J.~W.}\ \bibnamefont
  {Tukey}},\ }\href@noop {} {\enquote {\bibinfo {title} {The measurement of
  power spectra from the point of view of communications engineering},}\ }
  (\bibinfo {year} {1958})\BibitemShut {NoStop}%
\bibitem [{\citenamefont {Gaarde}\ \emph {et~al.}(2011)\citenamefont {Gaarde},
  \citenamefont {Buth}, \citenamefont {Tate},\ and\ \citenamefont
  {Schafer}}]{Gaarde-2011}%
  \BibitemOpen
  \bibfield  {author} {\bibinfo {author} {\bibfnamefont {M.~B.}\ \bibnamefont
  {Gaarde}}, \bibinfo {author} {\bibfnamefont {C.}~\bibnamefont {Buth}},
  \bibinfo {author} {\bibfnamefont {J.~L.}\ \bibnamefont {Tate}}, \ and\
  \bibinfo {author} {\bibfnamefont {K.~J.}\ \bibnamefont {Schafer}},\ }\href
  {\doibase 10.1103/PhysRevA.83.013419} {\bibfield  {journal} {\bibinfo
  {journal} {Phys. Rev. A}\ }\textbf {\bibinfo {volume} {83}},\ \bibinfo
  {pages} {013419} (\bibinfo {year} {2011})}\BibitemShut {NoStop}%
\bibitem [{\citenamefont {Gaarde}\ and\ \citenamefont
  {Schafer}(2013)}]{Mette-Chapter2}%
  \BibitemOpen
  \bibfield  {author} {\bibinfo {author} {\bibfnamefont {M.~B.}\ \bibnamefont
  {Gaarde}}\ and\ \bibinfo {author} {\bibfnamefont {K.~J.}\ \bibnamefont
  {Schafer}},\ }in\ \href@noop {} {\emph {\bibinfo {booktitle} {Attosecond
  Physics}}},\ \bibinfo {editor} {edited by\ \bibinfo {editor} {\bibfnamefont
  {L.}~\bibnamefont {Plaja}}, \bibinfo {editor} {\bibfnamefont
  {R.}~\bibnamefont {Torres}}, \ and\ \bibinfo {editor} {\bibfnamefont
  {A.}~\bibnamefont {Za{\"i}r}}}\ (\bibinfo  {publisher} {Springer Berlin
  Heidelberg},\ \bibinfo {year} {2013})\BibitemShut {NoStop}%
\bibitem [{\citenamefont {Freeman}\ \emph {et~al.}(1987)\citenamefont
  {Freeman}, \citenamefont {Bucksbaum}, \citenamefont {Milchberg},
  \citenamefont {Darack}, \citenamefont {Schumacher},\ and\ \citenamefont
  {Geusic}}]{Freeman-1987}%
  \BibitemOpen
  \bibfield  {author} {\bibinfo {author} {\bibfnamefont {R.~R.}\ \bibnamefont
  {Freeman}}, \bibinfo {author} {\bibfnamefont {P.~H.}\ \bibnamefont
  {Bucksbaum}}, \bibinfo {author} {\bibfnamefont {H.}~\bibnamefont
  {Milchberg}}, \bibinfo {author} {\bibfnamefont {S.}~\bibnamefont {Darack}},
  \bibinfo {author} {\bibfnamefont {D.}~\bibnamefont {Schumacher}}, \ and\
  \bibinfo {author} {\bibfnamefont {M.~E.}\ \bibnamefont {Geusic}},\ }\href
  {\doibase 10.1103/PhysRevLett.59.1092} {\bibfield  {journal} {\bibinfo
  {journal} {Phys. Rev. Lett.}\ }\textbf {\bibinfo {volume} {59}},\ \bibinfo
  {pages} {1092} (\bibinfo {year} {1987})}\BibitemShut {NoStop}%
\bibitem [{\citenamefont {Agostini}\ \emph {et~al.}(1989)\citenamefont
  {Agostini}, \citenamefont {Breger}, \citenamefont {L'Huillier}, \citenamefont
  {Muller}, \citenamefont {Petite}, \citenamefont {Antonetti},\ and\
  \citenamefont {Migus}}]{Agostini-1989-2}%
  \BibitemOpen
  \bibfield  {author} {\bibinfo {author} {\bibfnamefont {P.}~\bibnamefont
  {Agostini}}, \bibinfo {author} {\bibfnamefont {P.}~\bibnamefont {Breger}},
  \bibinfo {author} {\bibfnamefont {A.}~\bibnamefont {L'Huillier}}, \bibinfo
  {author} {\bibfnamefont {H.~G.}\ \bibnamefont {Muller}}, \bibinfo {author}
  {\bibfnamefont {G.}~\bibnamefont {Petite}}, \bibinfo {author} {\bibfnamefont
  {A.}~\bibnamefont {Antonetti}}, \ and\ \bibinfo {author} {\bibfnamefont
  {A.}~\bibnamefont {Migus}},\ }\href {\doibase 10.1103/PhysRevLett.63.2208}
  {\bibfield  {journal} {\bibinfo  {journal} {Phys. Rev. Lett.}\ }\textbf
  {\bibinfo {volume} {63}},\ \bibinfo {pages} {2208} (\bibinfo {year}
  {1989})}\BibitemShut {NoStop}%
\bibitem [{\citenamefont {Cohen-Tannoudji}(1996)}]{Cohen-1996}%
  \BibitemOpen
  \bibfield  {author} {\bibinfo {author} {\bibfnamefont {C.}~\bibnamefont
  {Cohen-Tannoudji}},\ }in\ \href {\doibase 10.1007/978-1-4612-2378-8_11}
  {\emph {\bibinfo {booktitle} {Amazing Light}}},\ \bibinfo {editor} {edited
  by\ \bibinfo {editor} {\bibfnamefont {R.}~\bibnamefont {Chiao}}}\ (\bibinfo
  {publisher} {Springer New York},\ \bibinfo {year} {1996})\ pp.\ \bibinfo
  {pages} {109--123}\BibitemShut {NoStop}%
\bibitem [{\citenamefont {Yakovlev}\ and\ \citenamefont
  {Scrinzi}(2003)}]{Scrinzi-2003}%
  \BibitemOpen
  \bibfield  {author} {\bibinfo {author} {\bibfnamefont {V.~S.}\ \bibnamefont
  {Yakovlev}}\ and\ \bibinfo {author} {\bibfnamefont {A.}~\bibnamefont
  {Scrinzi}},\ }\href {\doibase 10.1103/PhysRevLett.91.153901} {\bibfield
  {journal} {\bibinfo  {journal} {Phys. Rev. Lett.}\ }\textbf {\bibinfo
  {volume} {91}},\ \bibinfo {pages} {153901} (\bibinfo {year}
  {2003})}\BibitemShut {NoStop}%
\bibitem [{\citenamefont {Tate}\ \emph {et~al.}(2007)\citenamefont {Tate},
  \citenamefont {Auguste}, \citenamefont {Muller}, \citenamefont {Sali\`eres},
  \citenamefont {Agostini},\ and\ \citenamefont {DiMauro}}]{Tate-2007}%
  \BibitemOpen
  \bibfield  {author} {\bibinfo {author} {\bibfnamefont {J.}~\bibnamefont
  {Tate}}, \bibinfo {author} {\bibfnamefont {T.}~\bibnamefont {Auguste}},
  \bibinfo {author} {\bibfnamefont {H.~G.}\ \bibnamefont {Muller}}, \bibinfo
  {author} {\bibfnamefont {P.}~\bibnamefont {Sali\`eres}}, \bibinfo {author}
  {\bibfnamefont {P.}~\bibnamefont {Agostini}}, \ and\ \bibinfo {author}
  {\bibfnamefont {L.~F.}\ \bibnamefont {DiMauro}},\ }\href {\doibase
  10.1103/PhysRevLett.98.013901} {\bibfield  {journal} {\bibinfo  {journal}
  {Phys. Rev. Lett.}\ }\textbf {\bibinfo {volume} {98}},\ \bibinfo {pages}
  {013901} (\bibinfo {year} {2007})}\BibitemShut {NoStop}%
\bibitem [{\citenamefont {Beck}\ \emph {et~al.}(2014)\citenamefont {Beck},
  \citenamefont {Bernhardt}, \citenamefont {Warrick}, \citenamefont {Wu},
  \citenamefont {Chen}, \citenamefont {Gaarde}, \citenamefont {Schafer},
  \citenamefont {Neumark},\ and\ \citenamefont {Leone}}]{Leone-2014}%
  \BibitemOpen
  \bibfield  {author} {\bibinfo {author} {\bibfnamefont {A.~R.}\ \bibnamefont
  {Beck}}, \bibinfo {author} {\bibfnamefont {B.}~\bibnamefont {Bernhardt}},
  \bibinfo {author} {\bibfnamefont {E.~R.}\ \bibnamefont {Warrick}}, \bibinfo
  {author} {\bibfnamefont {M.}~\bibnamefont {Wu}}, \bibinfo {author}
  {\bibfnamefont {S.}~\bibnamefont {Chen}}, \bibinfo {author} {\bibfnamefont
  {M.~B.}\ \bibnamefont {Gaarde}}, \bibinfo {author} {\bibfnamefont {K.~J.}\
  \bibnamefont {Schafer}}, \bibinfo {author} {\bibfnamefont {D.~M.}\
  \bibnamefont {Neumark}}, \ and\ \bibinfo {author} {\bibfnamefont {S.~R.}\
  \bibnamefont {Leone}},\ }\href
  {http://stacks.iop.org/1367-2630/16/i=11/a=113016} {\bibfield  {journal}
  {\bibinfo  {journal} {New Journal of Physics}\ }\textbf {\bibinfo {volume}
  {16}},\ \bibinfo {pages} {113016} (\bibinfo {year} {2014})}\BibitemShut
  {NoStop}%
\bibitem [{\citenamefont {Yost}\ \emph {et~al.}(2009)\citenamefont {Yost},
  \citenamefont {Schibli}, \citenamefont {Ye}, \citenamefont {Tate},
  \citenamefont {Hostetter}, \citenamefont {Gaarde},\ and\ \citenamefont
  {Schafer}}]{Yost-2009}%
  \BibitemOpen
  \bibfield  {author} {\bibinfo {author} {\bibfnamefont {D.~C.}\ \bibnamefont
  {Yost}}, \bibinfo {author} {\bibfnamefont {T.~R.}\ \bibnamefont {Schibli}},
  \bibinfo {author} {\bibfnamefont {J.}~\bibnamefont {Ye}}, \bibinfo {author}
  {\bibfnamefont {J.~L.}\ \bibnamefont {Tate}}, \bibinfo {author}
  {\bibfnamefont {J.}~\bibnamefont {Hostetter}}, \bibinfo {author}
  {\bibfnamefont {M.~B.}\ \bibnamefont {Gaarde}}, \ and\ \bibinfo {author}
  {\bibfnamefont {K.~J.}\ \bibnamefont {Schafer}},\ }\href {\doibase
  10.1038/nphys1398} {\bibfield  {journal} {\bibinfo  {journal} {Nat Phys}\
  }\textbf {\bibinfo {volume} {5}},\ \bibinfo {pages} {815} (\bibinfo {year}
  {2009})}\BibitemShut {NoStop}%
\bibitem [{\citenamefont {Hostetter}\ \emph {et~al.}(2010)\citenamefont
  {Hostetter}, \citenamefont {Tate}, \citenamefont {Schafer},\ and\
  \citenamefont {Gaarde}}]{Tate-2010}%
  \BibitemOpen
  \bibfield  {author} {\bibinfo {author} {\bibfnamefont {J.~A.}\ \bibnamefont
  {Hostetter}}, \bibinfo {author} {\bibfnamefont {J.~L.}\ \bibnamefont {Tate}},
  \bibinfo {author} {\bibfnamefont {K.~J.}\ \bibnamefont {Schafer}}, \ and\
  \bibinfo {author} {\bibfnamefont {M.~B.}\ \bibnamefont {Gaarde}},\ }\href
  {\doibase 10.1103/PhysRevA.82.023401} {\bibfield  {journal} {\bibinfo
  {journal} {Phys. Rev. A}\ }\textbf {\bibinfo {volume} {82}},\ \bibinfo
  {pages} {023401} (\bibinfo {year} {2010})}\BibitemShut {NoStop}%
\bibitem [{\citenamefont {Soifer}\ \emph {et~al.}(2010)\citenamefont {Soifer},
  \citenamefont {Botheron}, \citenamefont {Shafir}, \citenamefont {Diner},
  \citenamefont {Raz}, \citenamefont {Bruner}, \citenamefont {Mairesse},
  \citenamefont {Pons},\ and\ \citenamefont {Dudovich}}]{Soifer-2010}%
  \BibitemOpen
  \bibfield  {author} {\bibinfo {author} {\bibfnamefont {H.}~\bibnamefont
  {Soifer}}, \bibinfo {author} {\bibfnamefont {P.}~\bibnamefont {Botheron}},
  \bibinfo {author} {\bibfnamefont {D.}~\bibnamefont {Shafir}}, \bibinfo
  {author} {\bibfnamefont {A.}~\bibnamefont {Diner}}, \bibinfo {author}
  {\bibfnamefont {O.}~\bibnamefont {Raz}}, \bibinfo {author} {\bibfnamefont
  {B.~D.}\ \bibnamefont {Bruner}}, \bibinfo {author} {\bibfnamefont
  {Y.}~\bibnamefont {Mairesse}}, \bibinfo {author} {\bibfnamefont
  {B.}~\bibnamefont {Pons}}, \ and\ \bibinfo {author} {\bibfnamefont
  {N.}~\bibnamefont {Dudovich}},\ }\href {\doibase
  10.1103/PhysRevLett.105.143904} {\bibfield  {journal} {\bibinfo  {journal}
  {Phys. Rev. Lett.}\ }\textbf {\bibinfo {volume} {105}},\ \bibinfo {pages}
  {143904} (\bibinfo {year} {2010})}\BibitemShut {NoStop}%
\bibitem [{\citenamefont {Botheron}\ and\ \citenamefont
  {Pons}(2010)}]{Botheron-2010}%
  \BibitemOpen
  \bibfield  {author} {\bibinfo {author} {\bibfnamefont {P.}~\bibnamefont
  {Botheron}}\ and\ \bibinfo {author} {\bibfnamefont {B.}~\bibnamefont
  {Pons}},\ }\href@noop {} {\bibfield  {journal} {\bibinfo  {journal} {Phys.
  Rev. A}\ }\textbf {\bibinfo {volume} {82}},\ \bibinfo {pages} {021404(R)}
  (\bibinfo {year} {2010})}\BibitemShut {NoStop}%
\bibitem [{\citenamefont {Doumy}\ \emph {et~al.}(2009)\citenamefont {Doumy},
  \citenamefont {Wheeler}, \citenamefont {Roedig}, \citenamefont {Chirla},
  \citenamefont {Agostini},\ and\ \citenamefont {DiMauro}}]{Doumy-2009}%
  \BibitemOpen
  \bibfield  {author} {\bibinfo {author} {\bibfnamefont {G.}~\bibnamefont
  {Doumy}}, \bibinfo {author} {\bibfnamefont {J.}~\bibnamefont {Wheeler}},
  \bibinfo {author} {\bibfnamefont {C.}~\bibnamefont {Roedig}}, \bibinfo
  {author} {\bibfnamefont {R.}~\bibnamefont {Chirla}}, \bibinfo {author}
  {\bibfnamefont {P.}~\bibnamefont {Agostini}}, \ and\ \bibinfo {author}
  {\bibfnamefont {L.~F.}\ \bibnamefont {DiMauro}},\ }\href {\doibase
  10.1103/PhysRevLett.102.093002} {\bibfield  {journal} {\bibinfo  {journal}
  {Phys. Rev. Lett.}\ }\textbf {\bibinfo {volume} {102}},\ \bibinfo {pages}
  {093002} (\bibinfo {year} {2009})}\BibitemShut {NoStop}%
\bibitem [{\citenamefont {Mauger}\ \emph {et~al.}(2012)\citenamefont {Mauger},
  \citenamefont {Kamor}, \citenamefont {Chandre},\ and\ \citenamefont
  {Uzer}}]{Mauger-2012}%
  \BibitemOpen
  \bibfield  {author} {\bibinfo {author} {\bibfnamefont {F.}~\bibnamefont
  {Mauger}}, \bibinfo {author} {\bibfnamefont {A.}~\bibnamefont {Kamor}},
  \bibinfo {author} {\bibfnamefont {C.}~\bibnamefont {Chandre}}, \ and\
  \bibinfo {author} {\bibfnamefont {T.}~\bibnamefont {Uzer}},\ }\href {\doibase
  10.1103/PhysRevLett.108.063001} {\bibfield  {journal} {\bibinfo  {journal}
  {Phys. Rev. Lett.}\ }\textbf {\bibinfo {volume} {108}},\ \bibinfo {pages}
  {063001} (\bibinfo {year} {2012})}\BibitemShut {NoStop}%
\end{thebibliography}%

\end{document}